**Miller Range 07687 and its place within the CM-CO clan**

Trygve Prestgard[1*], Lydie Bonal[1], Jolantha Eschrig[1], Jérôme Gattacceca[2], Corinne Sonzogni[2], Pierre Beck[1]

[1]Institut de Planétologie et d'Astrophysique de Grenoble, Université Grenoble Alpes, CNRS CNES, 38000 Grenoble (France)
[2] CNRS, Aix Marseille Univ, IRD, Coll France, INRA, CEREGE, Aix-en-Provence, France

* Corresponding author: trygve-johan.prestgard@univ-grenoble-alpes.fr.

*submitted to Meteoritics & Planetary Science*

**Abstract:**

Miller Range (MIL) 07687 is a peculiar carbonaceous chondrite officially classified as a CO3. However, it has been found to display unique petrographic properties that are atypical of this group. Moreover, Raman spectra of its polyaromatic carbonaceous matter does not reflect a structural order consistent with the metamorphic history of a type 3 chondrite. As a result, it has been suggested to be an ungrouped C2 chondrite with CO affinities, although it has not been fully excluded as a CO chondrite. The ambiguity of the meteorite's classification is the motivation behind the present study. We conclude that MIL 07687 is a unique carbonaceous chondrite with possible affinities to CO, CM and/or some ungrouped carbonaceous chondrites. The difficulty in classifying this meteorite stems from *(i)* its heavily weathered nature, which interferes with the interpretation of our oxygen (O-)isotopic measurements; *(ii)* the overlap in the petrographic and O-isotopic descriptions of various COs, CMs and ungrouped meteorites in the Meteoritical society database. Optical and infrared spectra are consistent with the meteorite's unequilibrated nature and indicate that it is probably mildly aqueously altered. Despite traces of aqueous alteration having previously been described in MIL 07687, this is the first time that the presence of hydrated amorphous silicates is reported. In fact, our results show that its present hydration is beyond that of most CO3s, less than most CM2s, and comparable to primitive CR2s. Consequently, we support the meteorite's C2-ung label, although a CO2 or CM2 classification cannot be fully excluded.

## 1. Introduction

Miller Range (MIL) 07687 is a small (5.5 g) carbonaceous chondrite discovered by the 2007 team of US Antarctic Search for Meteorites program. The meteorite was initially classified as a CO3 chondrite on the basis of its abundant and small chondrules (≤ 1 mm in diameter), presence of metal and sulfides occurring within and rimming the chondrules, and olivine and pyroxene elemental composition (initial petrographic description, available online on the Meteoritical Society database). However, since its initial classification, Brearley (2012) was the first to note that MIL 07687 differs significantly from typical CO3 chondrites due to various distinct features: (*i*) significantly higher matrix abundance (~ 68 vol. %: Brearley, 2012; and 63 vol. %: Vaccaro 2017); (*ii*) the lack of detectable fine-grained rims; (*iii*) the low abundance of refractory inclusions. Therefore, the meteorite is now considered (Antarctic meteorites classification database: https://curator.jsc.nasa.gov/antmet/samples/petdes.cfm?sample=MIL07687) as an ungrouped carbonaceous chondrite with affinities to the CO group, as also described by Haenecour et al.

(2020). Similarly to MIL 07687, there are various ungrouped chondrites, such as Acfer 094, Adelaide, and MacAlpine Hills (MAC) 88107, that appear to display some CO-like traits (e.g. chondrule size, matrix abundance and/or not very hydrated), although many of these have been considered meteorites with overlapping characteristics, showing possibly affinities with both the CO and CM chemical groups (Greenwood et al., 2019 and references therein). Unlike the latter meteorites, which have bulk oxygen (O)-isotopic measurements, to our knowledge, there appear to be no documented bulk O-isotopic measurements of MIL 07687. Consequently, its current classification appears to be mainly based on its petrography (Haenecour et al. 2020 and references therein).

In addition to its peculiar petrography, past research suggests MIL 07687 to be a highly primitive meteorite, notably by its apparently high concentration of amorphous material, $^{15}$N-enrichment and elevated abundance of presolar grains (Haenecour et al. 2020; McAdam et al., 2018; Vaccaro 2017). Indeed, amorphous silicates tend to be abundant in primitive chondrite matrices (Abreu and Brearley., 2010; Brearley, 1993; Dobrică and Brearley, 2020), and high bulk $^{15}$N content probably reflects material of strong interstellar heritage (Nakamura-Messenger et al., 2006). Furthermore, the $Cr_2O_3$ content in type-II chondrules of MIL 07687 are compatible with carbonaceous chondrites of petrographic type 3.0, and there is no evidence of zoning of forsteritic olivine (Brearley 2012), although Haenecour et al. (2020) report slightly lower $Cr_2O_3$ values than Davidson et al. (2014), possibly suggestive of minor thermal metamorphism.

Perhaps one of the most notable features about MIL 07687 is the presence of submillimeter-scale complex Fe-rich and Fe-poor matrix lithologies. The former regions are rich in fibrous assemblages of ferrihydrite, interpreted as being the result of partial (localized) aqueous alteration, possibly under highly oxidizing conditions (Brearley 2012, 2013, Haenecour et al., 2020). The oxidizing conditions are further supported by Ca-sulfates partially replacing Ca-carbonates (Brearley 2012, Haenecour et al., 2020). The Fe-rich regions appear to be separated from the primitive Mg-rich matrix by alteration fronts, rather than representing distinct brecciated lithologies. Interestingly, no phyllosilicates have been detected despite the evidence of aqueous alteration (Brearley, 2012, 2013, Haenecour et al. 2020). Brearley (2012) and Haenecour et al. (2020) further note that chondrule mesostasis occasionally appears to have suffered minor modifications in the regions that are in contact with the matrix. Consequently, MIL 07687 has been considered as a unique object, potentially providing the opportunity to study the earliest stages of aqueous alteration on primitive matrix material (Brearley, 2012). The identification and correct assessment of post-accretion processes has not only important consequences on our understanding of the geological processes themselves, but also on the initial properties and accretion conditions of chondrite parent bodies. Consequently, in order to provide a primary classification of MIL 07687, we decided to conduct bulk oxygen isotopic analysis. In regards to its secondary classification, we measured infrared (IR) spectra of bulk and matrix samples, along with bulk thermogravimetric analysis (TGA), in order to help provide the best possible constraints on the parent body processes the meteorite experienced. We additionally did an extensive search through the Meteoritical Society Database for unusual COs, CMs, and CO-CM clan members, that have similar petrographic descriptions with MIL 07687, and/or that have overlapping petrographic descriptions with the CO3 and CM2 chemical groups. Those with reported O-isotopic compositions are plotted in gray in Figs. 1 and 2.

## 2.1 Sample

A bulk sample (100 mg) of MIL 07687-5 has been provided by the NASA Meteorite Working Group, initially for the characterization of its metamorphic history through Raman spectroscopy of its polyaromatic carbonaceous matter (Bonal et al. 2016). As the thermal history of MIL 07687 appears to be different from that of the other CO chondrites considered in this study, we took advantage of the remaining sample to lead a further spectroscopic characterization and to characterize its degree of aqueous alteration.

The spectral data (of matrix fragments and bulk samples, in transmission and reflectance) of MIL 07687 are compared to those of CO, CR, CM, CV, CK, CI and ungrouped chondrites. The comparison data were either acquired in the present work or are from previous studies, as stated in Table 1. Most of the spectral data from previous studies are available online on the GhoSST/SSHADE spectral database (https://www.sshade.eu/db/ghosst: Schmitt et al., 2012).

We also performed TGA on a series of CO chondrites all listed in Table 2.

## 2.2 IR transmission spectra

*Bulk pellets:* a pellet made of KBr and bulk powdered MIL 07867 was prepared following the same procedure as described in Beck et al. (2014). Several mg of the bulk meteorite were ground in a dry agate mortar and out of this mass, 1.0 mg was weighed and mixed with 300 mg of commercial ultrapure KBr powder.

IR spectra were measured with a Brucker V70v spectrometer at the Institut de Planétologie et d'Astrophysique de Grenoble (IPAG, France). Spectra were acquired at 2 $cm^{-1}$ spectral resolution in the 5000 - 400 $cm^{-1}$ range under a primary vacuum (P = $10^{-3}$ mbar).

*Matrix fragments:* IR spectra were obtained with a BRUKER HYPERION 3000 infrared microscope (IPAG, France). The IR beam was focused through a 15x objective and the typical size of the spot on the sample was 40 x 40 $\mu m^2$. Spectra were measured at 4 $cm^{-1}$ spectral resolution with a MCT detector cooled with liquid nitrogen. Particular care was devoted to sample preparation, which is a critical issue in IR micro-spectroscopy. Samples must be thin (<20 $\mu m$) and their surfaces flat enough to avoid saturation of absorption and scattering artifacts, respectively (e.g., Raynal et al. 2000). Small matrix fragments (30 - 50 $\mu m$) were selected under a binocular microscope according to their color and texture. The matrix fragments were crushed between two diamond windows, allowing access to the 4000 - 650 $cm^{-1}$ spectral range. The diamond windows were loaded into an environmental cell, designed and built at IPAG. This cell enables temperatures up to 300 °C to be reached under primary or secondary dynamic vacuum (from $10^{-4}$ mbar down to $10^{-7}$ mbar). Optical access is permitted from both sides of the cell through ZnS windows, thus enabling measurements in transmission. Samples were progressively heated up to 250°C or 300°C, with typical steps at 100°C, 200°C, and spectra were recorded at each temperature step. The transmission spectra were automatically converted to absorbance (A = -log (T/$T_0$), where $T_0$ and T are the transmittance without and with the sample, respectively). To remove interferences and scattering effects, a spline baseline was subtracted from the raw data.

## 2.3 Reflectance spectra

The reflectance measurements of MIL 07687 were made by the SHADOWS instrument (IPAG, France, Potin et al., 2018). The method applied is that described by Eschrig et al. (2020): *(i)* we manually prepared 30 - 50 mg of an unsorted powder of sub-millimetre particle size using a mortar. Contrary to various other works (e.g. Cloutis et al., 2012a,b,c,d), we chose not to sieve the powder in order to better represent the grain-size heterogeneity expected of asteroidal regolith. *(ii)* The sample was secured in an environmental chamber (MIRAGE) under vacuum (P < 1 × $10^{-4}$ mbar) at ambient temperature. This was done in order to remove weakly bonded water molecules. The chamber was enclosed by a sapphire window, whose optical contribution was removed from the raw spectra thanks to an adapted algorithm. *(iii)* In regards to the observation geometry, we chose an incidence angle of i = 0°, and an emergence angle of e = 30°. The spectral resolution used varied from 0.048-$\mu m$ to 0.39-$\mu m$ (with a step of 0.02 $\mu m$). Note that the measurements were normalized to reference surfaces, more specifically by using a Spectralon[TM] (for wavelengths 0.4 – 2.1 $\mu m$) and an Infragold[TM] (2.1 – 4.2 $\mu m$).

## 2.4 TGA measurements

TGA was used to quantify the amount of hydrogen in the samples by measuring the mass loss upon heating. The first derivative (DTG) of the TGA curve enables us to determine the maximum peak temperature of each mass loss and gives an indication of the host mineral of volatile elements (see Fig. 1 in Garenne et al. 2014). Thermo-Gravimetric measurements were performed with a TGA-DSC3 Mettler-Toledo at Institut des Sciences de la Terre (ISTerre, Grenoble - France). A fragment of bulk sample was ground manually (around 50 mg in a mortar). 30 mg of this powder was extracted for TGA analysis and put inside a 150 μL alumina crucible under a 50 mL/min inert $N_2$ atmosphere. We included two separate samples of DOM 08006: -21 and -53. The former contained 30 mg, while the second was based on 50 mg of sample. This was done in order to study the mineralogical heterogeneity of a single meteorite. The mass loss from each sample was recorded as it was heated from 25°C to 1000°C with a heating rate of 10 °C/min. The TGA mass resolution is 1 μg, which corresponds to an absolute error of 0.07% for a total mass loss fraction of 5 %. The error on the temperature measurement is around 0.25 °C.

More specifically, the hydrogen in hydrated chondrites is mainly associated with phyllosilicates. In their work on type-2 chondrites, Garenne et al. (2014) showed that the 400 – 770°C TGA mass loss correlated well with the phyllosilicate content in CIs and CMs. King et al. (2015) found that mass loss due to serpentines took place between 200 - 800°C, through dehydration and dehydroxylation. In highly altered CMs, phyllosilicates appear to produce significant mass loss rates at ~400°C and 500 - 600°C (see Fig. 3 from Garenne et al., 2014). Carbonates were measured to decompose between 600 - 800°C by King et al., (2015), however mass loss due to carbonates in CMs has been observed to take place up to 900°C (Garenne et al., 2014). The breakdown temperatures of carbonates are variable depending on a variety of parameters, including chemistry, grain size and environment. Temperatures lower than 200°C correspond to the desorption of weakly bonded water (Garenne et al., 2014), while the 200 – 350°C region is dominated by the effects due to dehydroxylation of oxyhydroxides, most notably goethite (King et al., 2015).

Contrary to CIs and CMs, phyllosilicates appear to be of relatively low abundances, if not possibly absent, in MIL 07687 and most CO3s (Brearley 2006, 2012 and 2013, Haenecour et al. 2020). Consequently, we can expect the TGA mass loss in the 400 - 770°C interval of MIL 07687 and CO3s to be additionally (and significantly) affected by the decompositions of anhydrous minerals such as sulfides, magnetite and perhaps carbonates (Földvari 2011, King et al., 2015). Fe-sulfides are present in CO3s of all metamorphic grades (e.g. Alexander et al., 2018: ~2-3 vol%), and have also been reported in the matrix of MIL 07687 (~1 vol%: Vaccaro, 2017). In CO3s, sulfides do not seem to be the result of aqueous alteration (Brearley 2006). If we consider that MIL 07687, COs and CVs have comparable concentrations of sulfides (regardless of metamorphic grade), then, at most, we could possibly expect an additional ~0.4 wt.% mass loss to be associated with this mineral in the TGA 350°C – 700°C interval (Bonal et al., 2020).

Although absent in MIL 07687 (Vaccaro, 2017), magnetite is present to various degrees in primitive CO3s (2 – 8 vol%), and are generally absent in members of higher metamorphic grade (>3.2) (Rubin and Li, 2019). Minor amounts of magnetite however have been reported in Moss (>3.6) (Stokke, 2018). CO3s appear to be quite poor in carbonates (Brearley 2006), but these have been reported in MIL 07687 (Davidson et al., 2015).

Note that, while carbonates and magnetite are often considered result of aqueous alteration (e.g. Brearley 2006, Howard et al., 2014, Rubin and Lee, 2019), their abundances are difficult to determine via TGA. Under oxidizing condition, the latter is characterized by significant mass gain between 200 - 800°C (oxidation of magnetite into hematite: Földvari 2011), while the former is affected by a variety of factors, including grain size (e.g. King et al., 2015). However, as our experiments were done under inert conditions (probably no more than a few ppm of oxygen), we do not initially expect any mass gain due to magnetite to be highly significant. However, despite inert conditions, the presence of organics can probably lead to additional oxidation and/or reduction of certain minerals (e.g. magnetite and hematite), including at high temperature (e.g. Rudolph et al., 2012, Simmonds and Hayes, 2017), which may need to be considered when interpreting the TGA results.

Organics themselves (at least IOM) are expected to decompose progressively throughout our studied temperature range (Court and Sephton, 2014). Consequently, in CO3s, the mass loss associated with organics should vary between 0.1 - 0.5 wt.% (Alexander et al., 2007) within 25°C – 1000°C. However, due to the higher matrix abundance of MIL 07687, the mass loss contribution due to organics may be higher.

Consequently, despite the likely overlap from other minerals, we concluded that the 350 – 700°C range may be the best interval to quantify the phyllosilicate content (if significantly present) in MIL 07687 and CO3s, although we have nonetheless included the 400 – 770°C TGA values for the purpose of comparison with CMs. Moreover, as Fe-oxyhydroxides may also appear as parent alteration products in MIL 07687 and CO3s (Haenecours et al., 2020; Keller and Busseck, 1990), we have also chosen to specifically study the 200 - 700°C temperature interval. All values are summarized in Table 2.

In order to more accurately determine the composition of the released gas (i.e. the composition of the samples), the TGA curves were coupled with an Infrared Spectrograph (IRS). This is especially useful in regards to our samples, as various minerals (described above) may significantly contribute to the mass loss across the previously mentioned temperature ranges (which may hence strongly affect the interpretation of our results).

## 2.5 Oxygen isotopes

Measurement of oxygen isotopic compositions of a 1.5 mg aliquot of bulk powdered MIL 07687 was carried out at the Stable Isotopes Laboratory of CEREGE (Aix-en-Provence, France) using laser fluorination coupled with isotope ratio mass spectrometry (ir-MS) (see e.g., Alexandre et al., 2006; Suavet et al., 2010 for more details about the analytical procedure). The initial sample mass was 10 mg to ensure that measured aliquot is representative of the bulk meteorite. The three oxygen isotopic compositions were measured with a dual-inlet mass spectrometer Thermo-Finnigan Delta Plus. The oxygen isotope results are expressed in ‰ versus the international reference standard V-SMOW: $\delta^{18}O = [(^{18}O/^{16}O)_{sample}/(^{18}O/^{16}O)_{V-SMOW}-1] \times 1000$ and $\delta^{17}O = [(^{17}O/^{16}O)_{sample}/(^{17}O/^{16}O)_{V-SMOW}-1] \times 1000$. The $\delta^{18}O$ and $\delta^{17}O$ values of the reference gas were calibrated with measurements of NBS28 standard ($\delta^{18}O=9.60‰$, (Gröning 2004). $\Delta^{17}O$ is computed as $\Delta^{17}O = \ln(1+\delta^{17}O) - \lambda \times \ln(1+\delta^{18}O)$ with $\lambda=0.5247$ (Miller 2002). The $\delta^{17}O$ value of the NBS28 standard ($\delta^{17}O =5.026‰$) was computed so as to give $\Delta^{17}O=0‰$. The measurements of the session when MIL 07687 was measured were corrected on a daily basis using 1.5 mg quartz internal laboratory standard "Boulangé" (Alexandre et al. 2006; Suavet et al. 2010). The analytical uncertainties derived from long term repeated measurement (n= 59) of this internal laboratory standard are 0.076‰, 0.139‰, and 0.013‰ for $\delta^{17}O$, $\delta^{18}O$, and $\Delta^{17}O$, respectively.

## 3. Results

### 3.1. Oxygen isotopes of MIL 07687

We find that the bulk powder of MIL 07687 has an oxygen isotopic composition of $\delta^{18}O$ = -6.75‰ and $\delta^{17}O$ = -7.87‰. This is similar to bulk measurements of CO3 chondrites (Fig. 1), particularly those of some highly weathered Antarctic samples (Alexander et al. 2018): Its $\delta^{18}O$ values are slightly lower than typical COs (Figs. 1 and 2). Indeed, CO chondrites generally plot along and left of the CCAM line, and generally display negative bulk O-isotopic $\delta^{18}O$ values, following a fractionation line of ~0.6. Although, some COs listed in the Meteoritical Society Database appear to have positive $\delta^{18}O$ values (Fig. 1). CM chondrites tend to generally have a much heavier O-isotopic compositions ($\delta^{17}O > $ -2 ‰, $\delta^{18}O > + 4$ ‰), however many primitive CMs, and some CO/CM-like ungrouped meteorites are more $^{16}O$-rich, even plotting close to the CO footprint (Greenwood et al. 2019, Kimura et al. 2020, Lee et al. 2019, Meteoritical Society database: Fig 1 and 2). Anhydrous material from the Murchison CM2 meteorite also plot within the CO-field (Clayton and Mayeda, 1999: Figs. 1 and 2). In fact, MIL 07687 has a similar O-isotopic composition to EET 83355 (Figs. 1 and 2), a heated CM-like ungrouped carbonaceous chondrite (King et al., 2021). The O-isotopic composition of MIL 07687 is distinct from ungrouped CO/CM-like chondrites such as Acfer 094, Adelaide and MAC 88107, which display intermediate bulk isotopic properties of both typical CM and CO group members (Greenwood et al., 2019 and references therein).

### 3.2. IR spectra of matrix fragments

The IR spectra of MIL 07687 matrix fragments (Fig. 3) cover a spectral range that includes both the $SiO_4$ (Si-O) stretching (12500 – 800 cm$^{-1}$ or 8-12 μm, a.k.a. the 10-μm band) and the so-called 3-μm region. The latter is a complex region associated with various vibrational modes in phyllosilicates and oxyhydroxides (X–OH where X is a transition metal), as well as adsorbed and interfoliar molecular water. With the spectra having been measured at 250 - 300°C, oxyhydroxides and adsorbed molecular water are not expected to significantly contribute to the 3-μm band absorption. The peaks at 3.3 – 3.4 μm (~2900 cm$^{-1}$) are attributed to organic matter (C-H), and will not be discussed further in this work. We were able to obtain successful spectra of twelve matrix fragments. All display very common traits, as described below.

The spectral profiles of the $SiO_4$ (Si-O) stretching of MIL 07687 matrix fragments (Figs. 3 and 4) appear rounded and asymmetrical. More specifically, they appear to be composed of broad "bell"-shaped bands with a peak at ~9.8 μm (1020 cm$^{-1}$), alongside a weaker and narrower band at 11.2 μm (890 cm$^{-1}$). The Si-O profile for all matrix fragments measured in this work are very similar, although the prominence of the 11.2-μm band does vary slightly (being the strongest for fragments 1 and 2). Interestingly, the broad 9.8-μm band displays a similar morphology to that of saponite (Fig. 4), despite this mineral not having been previously described in this meteorite. The smooth and broad nature of the band is also indicative of silicates of low crystallographic order (Fu et al., 2017), hence consistent with amorphous phases detected by Vaccaro (2017) and McAdam et al. (2018).

The Si-O stretching region in MIL 07687 is similar to that observed in matrix fragments of some *(i)* CR2 chondrites (Bonal et al., 2013) (Fig. 4), *(ii)* CM2 chondrites (e.g. NWA 11588),

*(iii)* low petrologic type 3 CO chondrites DOM 08006 (CO3.00) and LAP 031117 (CO3.05) (Fig. 4). In regards to the 11.2-μm (890 cm$^{-1}$) band, its band location and narrow width are consistent with the presence of olivine.

The 3-μm band (measured at 250°C and 300°C, Fig. 4) is broad (2.7 – 3.5 μm or 3750 – 2900 cm$^{-1}$), rounded, asymmetric, with a peak absorption at 2.8 μm (3545 cm$^{-1}$). The profile is unlike those observed for CM2 chondrites (e.g. QUE 97990 and NWA 11588), but resemble those of primitive CR2 matrices (e.g. MET 00426 and QUE 99177). CM2s tend to have phyllosilicate-rich matrices (Brearley, 2006), while that of CR2s are generally dominated by hydrated amorphous silicates (Le Guillou et al., 2015). This is again consistent with the meteorite's amorphous silicate content, and with the lack of phyllosilicates detected in previous studies. MIL 07687 also seems to share similar 3-μm profiles to primitive CO3s (although much more intense in appearance). More details regarding the hydrated minerology of MIL 07687 will be led in Section 4.2.2.

We also observe absorption bands at *(i)* ~1600 cm$^{-1}$, consistent with water; *(ii)* 1379 cm$^{-1}$ and 841 cm$^{-1}$, consistent with carbonates; *(iii)* 1233 cm$^{-1}$, which might be due to evaporites or sulphates (Fig. 5). The intensity of each of these bands are highly variable between each fragment, with the 1233 and 841 cm$^{-1}$ bands being absent in most spectra.

### 3.3 IR spectra of bulk MIL 07687

Bulk measurement probes the combination of matrix, chondrules and CAI/AOA signatures. The IR spectrum of bulk MIL 07687 (Figs. 6 and 7) covers a spectral range that includes both the $SiO_4$ (Si-O) stretching (8-12 μm, a.k.a. 10-μm band) and bending (15 - 25 μm) regions, thus providing valuable information on the meteorite's bulk silicate composition. The spectrum is plotted alongside those of various type 2 carbonaceous chondrites (Fig. 6) and CO3s (Fig. 7), for comparison. Spectra of individual silicate minerals have also been added (from Salisbury, 1991). The spectra have been normalized according to the maximum absorbance of their 10-μm region.

We find that the 10-μm stretching band for bulk sample MIL 07687 displays various absorption peaks: *(i)* 9.3 μm, consistent with Mg-rich pyroxene; *(ii)* 11.2 μm, consistent with Mg-rich olivine; *(iii)* 9.9-μm: a feature that is consistent with the amorphous silicates observed in the IR spectra of matrix fragments (Fig. 3 and 4). The Si-O stretching profile is also especially similar to NWA 8631 (CO3.0) and EET 87522 (Heated CM2), probably indicating a similar silicate make-up between these meteorites.

In regards to the SiO₄ bending region (15 – 25 μm), MIL 07687 displays the presence of broad bands at 16.4 μm and 19.5 μm that are also consistent with the presence of Mg-rich pyroxene and olivine (e.g., Beck et al. 2014), the latter band likely being a combination of both minerals. A broad band appears around 23 - 24 μm, which likely represents a convolution between olivine, pyroxene and amorphous silicates (Figs. 6 and 7). In addition, the bulk spectra also display various faint absorption features between 13 μm and 16 μm, which are also observed in enstatite.

### 3.4 Reflectance spectra of bulk MIL 07687

The reflectance measurements (Fig. 8) cover a spectral range that extends from 0.4 μm to 4.0 μm. This includes the range commonly used for asteroid taxonomy (0.4 – 2.5 μm; e.g., DeMeo et al., 2009). This region is thought to be affected by a complex combination of numerous parameters, including the Fe content of various ferromagnesian silicates (e.g. olivine and phyllosilicates; Cloutis et al., 2011; Beck et al., 2018). The measured spectral range also includes the 3-μm region (Fig. 8) which is associated with various vibrational modes of phyllosilicates and oxyhydroxides (Metal-OH bonds), as well as adsorbed molecular water (see section 3.2).

The reflectance spectrum of MIL 07687 displays a red spectral slope (both in the visible and NIR), a broad 3-μm hydration band, and faint absorption features at 0.86 μm and 2.4 μm (Fig. 8). Otherwise, the spectrum is relatively featureless in the NIR. This spectrum is similar to the one previously published by McAdam et al. (2018). Poorly defined aliphatic bands are also detectable at 3.4 μm and 3.5 μm (Fig. 9).

With the exception of the visual slope, the 0.4-2.5 μm spectral region of MIL 07687 differs significantly from our CO3 chondrites: *(i)* the NIR spectral slope of MIL 07687 is redder than those generally observed in our spectra of CO3s; *(ii)* CO3 chondrites generally do not show obvious absorption features centered at 0.86 μm, but rather at 1.05 μm (with the exception of DOM 08006: narrow band at 0.98 μm) (Fig. 8; Eschrig et al., 2020). In addition, some CO3s (e.g. MIL 090785 and El Médano 389) also show broad bands at ~0.9-μm, attributable to oxyhydroxides (similarly to CR2 finds: Fig. 8). These meteorites are hence highly weathered, as supported by their strong visual slopes. The 0.86-μm feature of MIL 07687 is however significantly weaker than the 0.9-μm band observed in these weathered CO3s. The NIR reflectance of MIL 07687, and typical CO3s, also differ significantly from what we measured in our high-matrix abundance COs (NWA 8631 and NWA 11889). While the peculiar spectrum of NWA 11889 may be due to its high magnetite content (Meteoritical society database), it is not impossible that the ~1-µm feature in NWA 8631 is related to oxyhydroxides.

Although the 0.4-2.5 µm region of MIL 07687 differs from unheated CMs due to absence of a 0.7-µm phyllosilicate band (e.g. LEW 85311 and QUE 97990), it does however share a similar profile to some moderately heated (e.g. Stage II) CM2 chondrites: EET 87522 and MIL 090073 (Fig. 8). The phyllosilicate content in the latter two meteorites are likely deformed (amorphized) and dehydrated due to thermal metamorphism (Alexander et al., 2018; King et al., 2021). Overall, the NIR slope of MIL 07687 is comparable to CM chondrites. In addition to MIL 07687, absorption bands at 2.4 µm are also apparent in the spectra of some CMs, CRs and COs (Fig. 8 and Eschrig et al., 2020).

In regards to the 3-µm hydration band, the profile and intensity are similar to that of CR2 finds and highly weathered CO3s (Fig. 9), hence indicative of oxyhydroxides rather than phyllosilicates, consistent with matrix transmission spectra (section 3.2 and 3.3).

## 3.5 TGA of MIL 07687 and CO3 chondrites

The TGA curve reflects the loss of volatiles from the decomposition of minerals through increasing temperature, as described in section 2.2. Consequently, the composition of the meteorite will affect the mass loss curve. The DTG is used to better visualize specific temperatures of anomalously high mass loss rates. The TGA mass loss values are presented in Table 2, and the associated curves are displayed in Figs. 10 and 11.

In the case of MIL 07687, high rates of mass loss are measured between at 400 – 500°C and 650 – 800°C. We further note that two distinct "peaks" in the DTG curve appear both within the former (400°C and 440°C) and latter region (670°C and 765°C). Below 350°C, we see signatures at 210°C due to Ferrihydrite (Fig. 13), consistent with its detection in previous studies (e.g. Haenecour et al. (2020). The narrow DTG peak at 130°C may be associated with evaporites (gypsum), if not molecular water. This would be consistent the meteorite's description in the Meteoritical Society Database, and possibly with our matrix spectra (section 3.1 and Fig. 5).

The overall mass loss of MIL 07687 is significantly higher than those measured for the CO3s; this statement being applicable to all predefined temperature ranges (Table 2). We note that the 400 – 770°C mass loss values for MIL 07687 are similar to those of poorly altered CM2 chondrites, such as LEW 85311 and QUE 97990 (Garenne et al., 2014). Moreover, when normalized according to their matrix abundance, the 200 - 700°C, 350 - 700°C and 400 – 770°C mass loss values for MIL 07687 remain superior to all CO3s considered (based on the matrix abundances in Table 2).

Although our CO3s present lower mass loss values (Table 2), the morphology of most CO3 TGA/DTG curves present broad similarities to that of MIL 07687 (Figs. 10 and 11): *(i)* All samples, with the exception of Moss (which presents no obvious mass loss, even a very mild mass gain), show an especially rapid mass loss between 650 – 900°C. *(ii)* Furthermore, DOM 03238, MIL 05104, MIL 07193 and ALHA 77003 also show a similar mass loss feature between 400 – 500°C (although less pronounced than in MIL 07687). These features are particularly interesting in the case of ALHA 77003, as it has a similar metamorphic grade to Moss (>3.6). A small "inflexion" in the TGA mass loss curve of Moss is apparent at ~890°C.

We note obvious similarities between the two DOM 08006 fragments, both in terms of mass loss values and TGA/DTG curve morphology (Fig. 10). We also observe an overall strong similarity between MIL 05104, MIL 05024, and MIL 07193, which are all probably paired (Meteoritical society database, Alexander et al., 2018), which is further supported by their very similar metamorphic grades (Bonal et al., 2016). However, small differences appear: *(i)* MIL 05024 lacks a 400-500°C "peak" in its DTG curve *(ii)* MIL 07193 displays a distinct mass loss trend between 650 – 900°C compared to MIL 05104 and MIL 05024. MIL 07193 also has more mass loss associated with ferrihydrite than the other two.

Lastly, in the IRS spectra of all our samples, we observe the presence of two neighboring bands at 4.6 and 4.75 μm, consistent with carbon monoxide, indicative of carbonates (see section 4.2.2.). They appear between ~600 and ~950°C (Figs. 10 and 11), and are strongest in the case of MIL 07687 (Figs. 12 and 13). The peak intensity of the bands coincide with the peak of the 650-900°C mass loss, as well as the "inflexion" at 890°C in the TGA curve of Moss. The intensity of the bands is similar between primitive (3.0 - 3.1) CO3s (30 mg samples), with the exception of DOM 03238 in which they appear significantly stronger. The intensity is also stronger for the 50 mg DOM 08006-53 sample, which is simply due to its higher mass. Those of ALHA 77003 and Moss are the weakest. The carbon monoxide bands in the DOM 08006 samples appear to have two main peaks of intensity (~650°C and ~790°C: Fig. 12), also consistent with their TGA/DTG curves.

Unfortunately, no absorption band in the IRS spectra were found to match that of the 400 – 500°C feature. We do not believe it to be associated with Fe-sulfides, as Moss does not show this mass loss feature. However, as noted earlier, the 400 – 500°C DTG peak is absent in DOM 08006, MIL 05024 and Moss. These all appear to be relatively poorly weathered (both the former have a weathering grade of A/B, and Moss is a recent fall). DOM 08006 and Moss also lack obvious resolvable TGA/DTG signatures of ferrihydrite and other oxyhydroxides (Fig. 13). The feature is most obvious in MIL 07687, DOM 03238 and ALHA 77003, two of displayed a noticeably "rusty" appearance under optical polarizing light (ALHA 77003 and MIL 07687). This was however not the case with DOM 03238, where optical observations suggested the extent terrestrial weathering to be less severe. However, evaporites may be present based on the narrow DTG peak at 120°C, which could indicate significant terrestrial weathering. The absence of a 400 – 500°C feature in MIL 05024, despite being present in MIL 07193 and MIL 05104, is also consistent with a terrestrial origin of the host mineral.

Moss (a fall), of similar metamorphic grade to ALHA 77003 (a find, significantly weathered), does not display any significant mass loss. This might suggest that the TGA mass loss in ALHA 77003 may be due to terrestrial weathering (assuming no post-metamorphic hydration). Moreover, seen that all CO3 finds, and MIL 07687, display similar TGA/DTG profile, it could be that their TGA mass loss values too are strongly influenced by terrestrial weathering products. However, despite this, we note that the matrix-normalized TGA 200 – 700°C and 350 – 700°C mass loss decreases with metamorphic grade (Table 2).

## 4 Discussion

### 4.1. MIL 06787 is a weathered chondrite that is difficult to classify

As described in section 3.1. MIL 07687 was found to have a bulk O-isotopic composition that is most consistent with CO chondrites. In fact, its O-isotopic composition is within the lighter isotopic range of the CO group. At first glance, this could indicate that MIL 07687 is indeed a true CO chondrite. In fact, three meteorites listed in the Meteoritical Society database appear to have comparable matrix abundances to MIL 07687: Catalina 008 (58%), Mdaouer (62%) and NWA 11889 (61%). Both Catalina 008 and NWA 11889 have O-isotopic measurements consistent with the CO group. Interestingly, over a dozen CO3 entries are listed as having a matrix abundance of at least ~50% (e.g. NWA 8631: 53%). Moreover, Davidson et al. (2019) interestingly noted that DOM 08006 is very poor in refractory inclusions (1%), especially AOAs, hence again consistent with MIL 07687. The Meteoritical bulletin also lists NWA11559 (CO>3.6) as not containing any detectable CAIs. We also measured the average chondrule size of MIL 07687 as being a close match to the CO chondrite average (Table 3). Lastly, our MIR spectrum of NWA 8631 (CO3.0) is similar to MIL 07687, indicating a similar bulk silicate content. Although these observations may support MIL 07687 as having affinities with the CO group, they are probably not sufficient to classify it as a true member. Indeed, there are several points to consider.

Firstly, not only does MIL 07687 share petrographic similarities with these unusual CO3s, but also with the CM group and several ungrouped CM-CO clan members. In fact, the boundaries between all three taxonomic classes appear to be quite blurry. In average, CM chondrites have a matrix abundance of 70% (50 – 90%, based on values from the Meteoritical Society Database), fewer refractory inclusions and similar chondrule sizes to CO chondrites (Table 3). They are hence, in average, petrographically more alike MIL 07687 than "typical" CO3s (Table 3). Moreover, the extent of the O-isotopic composition of the CM group has also been brought into question. Traditionally (other than their bulk elemental composition, and average petrographic properties: Krot et al., 2014, Weisberg et al., 2006), CM chondrites are distinguished from COs by their level of hydration (CMs being significantly more hydrated than COs) and by their O-isotopic composition (gap between $\delta^{18}O= 0$ and $+ 4$ ‰: Greenwood et al., 2019 and references therein). However, this view has been challenged by the discovery of several mildly altered CM/CO-like objects that populate the O-isotopic "gap" defined by "typical" COs and CMs (e.g. Greenwood et al., 2019), many of which may actually be primitive CMs (Kimura et al., 2020). The generally heavier O-isotopic composition of CMs (relative to COs) may be partially attributed to aqueous alteration (interaction with $^{16}O$-poor fluids), with the O-isotopic start-point of CM material being similar to that of bulk COs. This is supported by the CO-like O-isotopic composition of anhydrous CM material and the $^{16}O$-rich composition of some highly primitive CMs (Fig. 1 and 2., Clayton and Mayeda, 1999, Greenwood et al., 2014, Kimura et al., 2020). In fact, Lee et al. (2019) measured the mildly altered CM LEW 85311 to have an O-isotopic composition similar to that of CO falls (Fig. 1 and 2). The extent of the CM field towards lighter O-isotopic compositions is thus not well defined and may potentially extend into the CO field.

We should also note that some CM chondrites have average chondrule sizes that are closer to MIL 07687 than the CM-average (e.g. LEW 85311 and Winchcombe: Choe et al. (2010) and Meteoritical Society database). One could thus ask if MIL 07687 actually has closer affinities with the CM group. Moreover, although MIL 07687 appears to have a significantly higher $^{18}O$ content than CMs, it is likely that its intrinsic O-isotopic composition was closer to the CCAM line prior to terrestrial residency (see below). However, fine-grained rims do seem to be common in CMs (Meteoritical Society database), which is unlike MIL 07687.

Lastly, the Meteoritical Society Database lists a large variety of ungrouped chondrites with properties that are either intermediate, or overlap, with both the CO and CM groups (and MIL 07687). Like MIL 07687, several ungrouped chondrites appear to have O-isotopic compositions that are close or consistent with that of the CO group. Some appear to be hydrated (e.g. EET 83226/83355 and NWA 13249, with somewhat similar petrographic descriptions to CM2s in the Meteoritical bulletin) while others appear relatively anhydrous (e.g. El Médano 200 and NWA 12957). MIL 07687 has a comparable matrix abundance and chondrule size to several of these (see Table 3). Consequently, MIL 07687 might possibly have closer affinities with some of these meteorites, despite its lighter O-isotopic composition (which might be due to terrestrial weathering, see below).

Based on the oxygen isotopic composition and petrographic diversity reported for CM-CO members in the Meteoritical society database (commonly used for classification), the boundaries between what we consider CO, CM and ungrouped chondrites appears to be quite blurry (Table 3, Figs. 1 and 2). Due to the overlaps between these groups, it is clear that MIL 07687 cannot be attributed a taxonomic classification simply based on its petrography and O-isotopic composition.

In addition, terrestrial weathering poses a problem when interpreting the O-isotopic composition. Indeed, as noted in section 3.1, although the O-isotopic composition of MIL 07687 is most consistent with the CO group, it appears slightly more $^{18}$O-poor than "typical" members, a feature particularly observed by Alexander et al. (2018) in highly weathered CO3 chondrites (e.g. MIL 03442 and MIL 090785). Their work concludes that this "offset" from the typical CO trend is not representative of the original O-isotopic composition of these meteorites, but instead results from the contribution of Antarctic terrestrial weathering products. More specifically, the bulk isotopic composition of CO finds may evolve as a result of progressive interaction with terrestrial water, leading to a $\delta^{17}$O and $\delta^{18}$O values that shifts towards the Terrestrial Fractionation Line (TFL), in vicinity of the Standard Light Antarctic Precipitation (SLAP) in the case of Antarctic finds (Alexander et al., 2018). Indeed, this can lead to heavily weathered Antarctic COs displaying relatively light $^{17}$O and $^{18}$O compositions compared to "typical" COs. An initial analysis by the Johnson Space Center (JSC) found that MIL 07687 displays a black/brown fusion crust and matrix, including evaporates and the presence of "rusty" chondrules (initial petrographic description, available online on the Meteoritical Society database), leading to the attribution of a "Ce" weathering grade (i.e. strongly weathered, containing evaporates).

In addition to many of the TGA features mentioned in section 3.5, the intensely weathered nature of MIL 07687 is likely supported by various features in our results, including: *(i)* the TGA mass loss for temperatures lower than 200°C *(ii)* the 3-μm band and *(iii)* the optical slope in the reflectance spectrum. The two former parameters are strongly influenced by adsorbed molecular $H_2O$ (although the former also possibly includes the contribution due to evaporates at 130°C, i.e. gypsum, a weathering product). Indeed, Bonal et al. (2020) found that low temperature mass loss was a reliable indicator of terrestrial weathering in CV chondrite samples. More specifically, they found that highly weathered CVs were measured to have significantly larger TGA < 200°C mass loss values (between 1.2 to 3.5 wt.%, compared to < 1 wt.% for most CVs), hence reflecting strong (or long-term) exposure to terrestrial water. The absorption feature at 0.86 µm may also represent an excess of oxyhydroxides due to terrestrial weathering, similar to the feature observed in the highly weathered CO3s MIL 090785 and El Médano 389 at ~0.9 µm (Fig. 8), and in CR2s (Cloutis et al 2012a, Fig. 8).

McAdam et al. (2018) also suggested that the red NIR slope observed in MIL 07687 due to the influence of terrestrial weathering products, however Eschrig et al. (2020) found that weathered COs tend to have bluer NIR slopes (with poorly weathered COs and falls being red-sloped). Moreover, the weathered CO3 samples presented in Fig. 8 all seem to have neutral or blue NIR slopes. This raises the question of whether the NIR slope may actually be an intrinsic spectral feature of the meteorite's parent body, rather than an influence of weathering products. An in-depth discussion regarding the origin of this feature will be described in section 4.2.2.

In order to have a better idea of the pre-weathered O-isotopic composition of MIL 07687, an estimation was made based on the Lever's rule. If we consider oxyhydroxides as the main terrestrial weathering product, we can conclude, based on Lever's rule, that MIL 07687's original unweathered composition would possibly fall within the typical range of CO chondrites, closest to Felix, if it was to plot on the CCAM line. Although it is difficult to provide an estimation of the $\delta^{17}O$ and $\delta^{18}O$ values, our projection suggests that the original O-isotopic composition was unlike that of "typical" CM chondrites. We suggest that MIL 07687 should remain classified as "ungrouped" until further evidence is provided on its nature.

## 4.2 Secondary history of MIL 07687
### 4.2.1 MIL 07687 is a poorly metamorphosed CM-CO clan member

The $SiO_4$ (Si-O) stretching (~10 µm) and bending (15 – 25 µm) regions of MIL 07687 are dominated by the presence of Mg-rich olivine, pyroxene, and amorphous silicates (see section 3.3; Figs. 6 and 7). The lack of a 1.05-µm band in the reflectance spectrum of MIL 07687 (Figs. 8) is also consistent with the dominant Mg-rich mineralogy of its olivine. It is believed that the absorption band at 1.05 µm most likely reflects $Fe^{2+}$ crystal field transitions of olivine, while the band at 2 µm is partially due to $Fe^{2+}$ crystal field transitions of spinel (Cloutis et al., 2004; Cloutis et al., 2012b, c). Consequently, chondrites of higher FeO-rich olivine and spinel concentrations will display stronger 1.05-µm and 2-µm absorption bands.

This is most notably the case of type >3.1 COs (e.g. ALHA 77003 and El Médano 389) and type 2 heated stage IV chondrites (e.g. PCA 02010) (Fig. 8), suggesting that MIL 07687 has hence been heated less than these meteorites.

Although the bulk Si-O band profile of MIL 07687 has similarities with some primitive chondrites (e.g. CO3.0 NWA 8631), its reflectance and bulk IR transmission spectra resemble that of the stage II heated CM2 EET 87522. This is mainly because olivine and amorphous silicates can be significantly present in both primitive and Stage II/III heated chondrites (e.g. Abreu and Brearley, 2010; King et al., 2021). With increasing temperatures these poorly-crystalline silicates (regardless of being primary or secondary in origin) are progressively crystallized into purely anhydrous minerals, mainly olivine (e.g. King et al., 2021), hence the higher olivine abundance in CO>3.0 and Stage III/IV chondrites (Figs. 6 and 7). This can also clearly be observed in our IR spectra of CO3 matrix fragments, where the primary amorphous silicates decrease in abundance, relative to olivine, in the order of DOM 08006 (3.00), LAP 031117 (3.05) and DOM 03238 (3.1) (Fig. 4).

Our matrix and bulk IR spectra show that the abundance of amorphous silicates in the matrix of MIL 07687 are comparable to that in primitive CR2 matrices, moderately heated CM2s, and more than in CO3s. Although this could indicate that it has suffered a type-2 thermal history, our IR spectra alone cannot say whether MIL 07687 is highly primitive or underwent mild-to-moderate short-term thermal metamorphism.

Doubt in regards to the interpretation of the thermal history is removed thanks to Raman spectroscopy of polyaromatic carbonaceous matter. Indeed, the Raman spectra acquired on matrix grains of MIL 07687 (Bonal et al., 2016) are characterized by a high fluorescence background and by a shallow spectral valley between the Raman D- and G-bands (Fig. 14a). This is typical of type 2 chondrites, and this simple visual inspection allows to distinguish MIL 07687 from primitive CO3 chondrites, such as DOM 08006 (3.0; Bonal et al., 2016). To characterize the thermal history of type 2 chondrites, the most sensitive spectral parameters are the ones related to the G-band: its Full Width at Half Maximum and its position (e.g., Quirico et al., 2018). In particular, Quirico et al. (2018) distinguished 3 Raman groups among 40 CM and ungrouped chondrites. The R1 group contains chondrites with no detectable structural modifications by heating. R3 contains chondrites with slight structural modifications induced by a weak heating. The R2 group is characterized by Raman parameters pointing to a higher degree of structural order than R1 and R3, unambiguously reflecting thermal events. The Raman parameters describing the spectra of MIL 07687 are plotting among "R1" type 2 chondrites (Fig. 14b). This clearly shows that MIL 07687 escaped any significant thermal metamorphism.

**4.2.2 MIL 07687 likely experienced an episode of aqueous alteration**

As underlined in section 3.2., the detection of a 3-μm band (broad with a peak intensity at 2.8 μm; Figs. 5 and 6) shows that (at least part) of the matrix amorphous silicates in MIL 07687 are hydrated. This suggests that the meteorite's matrix possibly suffered incomplete and incipient aqueous alteration of its matrix, perhaps similarly to primitive CR2s (Le Guillou et al., 2015). While $Fe^{3+}$oxyhydroxides possibly produced by aqueous alteration have been detected in MIL 07687 (Brearley 2012, 2013; Haenecour et al., 2020), this appears to be the first time, to our knowledge, that hydrated amorphous silicates have been reported in this meteorite. The question is whether these hydrated minerals are of terrestrial or parent-body origin.

As established in section 4.1., MIL 07687 is a highly weathered meteorite. However, if these minerals were to solely be of terrestrial origin, this would suggest that MIL 07687 would have been an extremely pristine meteorite prior the onset of Antarctic weathering (thus very rare). Moreover, the distribution of hydrated amorphous silicates in MIL 07687 seems quite homogenous throughout the sample, which would unlikely be the case is they were solely due to terrestrial weathering.

The next question is whether our TGA data can also be used as a tracer of the meteorite's hydration. At first glance, according to our initial definition (section 2.4), the high-temperature TGA 350-700°C mass loss (5.8 wt.%) would indicate phyllosilicates being present in this meteorite. However, as we have observed in our results (section 3.5, Fig. 13), we must note the overlap with other minerals such as possibly oxyhydroxides at around 400 – 500°C, carbonates that have an onset at around 650°C, and organics that may decompose throughout the temperature ranges studied. In fact, MIL 07687 and all CO3 finds have similar TGA curves, suggesting that many of the minerals significantly contributing to the TGA mass loss of these meteorites may actually be of terrestrial origin. Consequently, parent-body phyllosilicates (if present) are unlikely to significantly contributing to the mass loss, consistent with our IR spectra.

However, we do observe a possible trend: *(i)* the matrix-normalized TGA 200 – 700°C and 350-700°C mass loss decreases with increasing thermal metamorphism (see Table 2); *(ii)* the intensity of the carbon monoxide band is weaker in highly metamorphosed CO3s compared to MIL 07687 and primitive CO3s (Fig. 14 and 15); *(iii)* although weak, possible signatures of carbonates are present in the TGA/IRS data of Moss, despite being a fall. Thus, in addition to hydrated amorphous silicates, our data possibly suggests that a subset of oxyhydroxides and carbonates are also parent body alteration products in CO3s and MIL 07687. In the case of oxyhydroxides, "dendritic" ferrihydrite in the matrix of MIL 07687 is already regarded as a possible aqueous alteration product (e.g. Haenecour et al., 2020). Moreover, oxyhydroxides have also previously been mentioned as alteration products in COs and CMs (Keller and Buseck, 1990; Pignatelli et al., 2016), as have carbonates in CMs (e.g. De Leuw et al., 2010).

Alternatively, carbon monoxide emissions in our samples may also result from the interaction between organic matter and magnetite at high temperatures (e.g. Rudolph et al., 2012). Both organic matter and magnetite are more commonly observed in primitive versus more metamorphosed CO3s (Alexander et al., 2007; Rubin and Li, 2019). This would explain the stronger carbon monoxide bands observed in DOM 03238 (a magnetite-enriched CO3.1) versus the other COs. Vaccaro et al. (2017) did not detect magnetite in the matrix of MIL 07687, which would suggest that it is significantly richer in carbonates than the CO3s studied in this work. Both alternatives support the detection of aqueous alteration in MIL 07687 and CO3s through TGA and a larger number of CO samples studied by TGA should be considered for future reference.

Although the TGA mass loss may be difficult to use in order to quantify the hydration of MIL 07687 relative to CO3s, the IR transmission spectra of MIL 07687 matrix fragments support this meteorite as being more hydrated than our CO3s, and comparably so to some CR2s. As previously mentioned, the MIR matrix spectral profile of MIL 07687 is very similar to primitive CR2 chondrites, both in terms of their average 3-μm and 10-μm profile and intensity, indicating that they contain a similar abundance of Fe-rich amorphous hydrated silicates in their matrices. However, phyllosilicates have been reported in pristine CR2s (Harju et al., 2014; Le Guillou et al., 2015), unlike MIL 07687, indicating that they CR2 may be slightly more aqueously altered. Consequently, based on the comparison of the proposed CM petrographic classification by Kimura et al. (2020), and that by Harju et al. (2014) for primitive CR2s, we could estimate MIL 07687 as having a petrographic grade of 2.9. This would further be supported by the type-2 thermal history of this meteorite. We note however that hydrated amorphous silicates are not detected in the bulk IR transmission spectra of primitive CR2s (Fig.7), which is most likely due to their lower matrix abundances than in MIL 07687.

MIL 07687 is less hydrated than the relatively mildly altered CM2.4-2.7s as shown by its lack of phyllosilicate absorption bands in its bulk and/or matrix IR spectra (Figs. 4 and 7). This is despite the comparable TGA 400 – 770°C values between MIL 07687, LEW 85311 (CM2.7) and QUE 97990 (CM2.6), thus showing that this temperature range is not reliable when quantifying the hydration. This is likely due to the difficulty in quantifying the mass loss due to carbonates and/or magnetite.

Lastly, we propose that the red reflectance NIR slope reflects the content of hydrated amorphous silicates present in MIL 07687. Indeed, we observe similar slopes in the spectra of MIL 090073, EET 83355 and EET 87522 (Fig. 8), all which are known to contain deformed (amorphized) phyllosilicates (Alexander et al., 2018, King et al., 2021), rather than crystalline phyllosilicates. For comparison, LEW 85311 and QUE 97990 also displays a similar red NIR slopes, but are accompanied by distinct 0.7-µm and 2.8-µm features (indicative of significant crystalline phyllosilicate content). We can rule out metal as the likely cause for their red slopes (e.g. in CR2s: Cloutis et al. 2012a.), as they are rare in these meteorites (less than 1%), MIL 07687 included (King et al., 2021, Vaccaro, 2017).

### 4.3. What if MIL 07687 is a CO2 chondrite?

If we were to identify MIL 07687 as a CO2.9 chondrite, we would extend the known petrographic range of the CO group, and thus provide new constraints on the geological processes occurring on their parent body (assuming that our currently recognized CO chondrites correctly sample one single parent body). The final hydration state of a sample (i.e. presence of hydrated amorphous silicates) is controlled by the interplay of heat and water. On the CO parent body, the heat source was likely of radiogenic origin (i.e. long-term metamorphism). If indeed a CO2 chondrite, MIL 07687 would possibly have originated from a location within the parent asteroid that escaped the extent of thermal metamorphism experienced by CO3s, while its relatively hydrated nature would show that it better retained (and/or interacted with) the surrounding fluids (or did not experienced dehydration thanks to a lower peak metamorphic temperature). This can be explained if MIL 07687 represents a portion of the parent body that accreted a higher abundance of ice. More specifically, the higher abundance of ice may have locally buffered the peak metamorphic temperatures (Grimm and McSween, 1989) and also promote a more extensive interaction of fluid as well as the conservation of hydrated minerals.Consequently, if it was to be a CO chondrite, MIL 07687 may be evidence that the CO parent body was heterogeneous, possibly having accreted variable amounts of water ice that was heterogeneously distributed throughout the parent asteroid.

Another possible explanation would be that MIL 07687 formed within a relatively hydrated region (or layer) close to the parent body surface (perhaps it would have accreted onto the parent body later than most CO material). Indeed, nearly all the CO3s in the Meteoritical

Society Database (2021) with comparable matrix abundances to MIL 07687 (~50% or more) all appear to be quite primitive (often listed as CO3.0s), with similar $Cr_2O_3$ values to MIL 07687. Thus, if these meteorites are indeed COs (which needs to be verified), and assuming that this meteorite group only samples one single primary parent body, this would possibly suggest that MIL 07687, along with these meteorites, would represent a poorly heated, matrix-enriched, and potentially hydrated, superficial layer of the CO parent body. Unfortunately, to our knowledge, we were not able to find any noble gas data in the literature that may give insight into whether or not MIL 07687 was located near the surface of its parent body. In both cases, not only would it imply that the parent body was petrographically heterogeneously, but it would further suggest that the overall water content on the CO parent body would likely have been higher than previously considered. The idea of an unevenly hydrated CO parent planetesimal may be supported by the discovery of unheated hydrated clasts in NWA 1232, a brecciated CO3 (Matsumoto et al. 2015), assuming that these are not xenoliths. There are also several COs that appear to contain phyllosilicates (e.g. Acfer 374, EET 90043 and NWA 11790: Meteoritical society database).

Although it is currently not possible to conclude whether MIL 07687 actually belongs to the CO group, it is an interesting path to explore as it has important implications on our understanding of their parent body.

## 5. Conclusion

Based on a combination of bulk oxygen isotopic measurements, spectroscopy and thermogravimetric analysis studies, we are able to conclude that:

1. MIL 07687 cannot be provided a primary classification by its O-isotopic composition and petrography alone, due its overlapping petrographic descriptions and bulk O-isotopic composition with various CO, CM and ungrouped chondrites in Meteoritical Society database. Significant terrestrial weathering further complicates the interpretation of the O-isotopic composition. We recommend that this meteorite remains classified as "ungrouped", although a CO or CM classification cannot be excluded. We do estimate however that the $\delta^{17}O$ and $\delta^{18}O$ values were lower than Acfer 094, Adelaide, MAC 88107 and most CM2s (more alike COs and some ungrouped CCs).

2. MIL 07687 experienced a thermal history similar to type 2 chondrites, as attested by the structural order of the polyaromatic matter characterized by Raman spectroscopy.

3. MIL 07687 has likely experienced more extensive aqueous alteration than most CO3 chondrites, and less than most known CM2s. The hydration probably resulted in hydrated Fe-rich amorphous silicates, oxyhydroxides, and possibly carbonates. Nevertheless, the alteration was likely very mild and did not include the entirety of its matrix. Significant nucleation of phyllosilicates did not likely occur. The extent of aqueous alteration is comparable to that observed in poorly hydrated CR2s. The red NIR slope of MIL 07687 might be a result of its hydrated amorphous silicate content. We recommend a petrographic grade of about 2.9, if these hydrated minerals are indeed pre-terrestrial.

4. We explore the hypothetical implications of MIL 07687 if it was to be a CO2 chondrite. If true, it would suggest that the CO parent body was geologically heterogeneous, assuming that all COs sample a single parent asteroid. It would further suggest that water was unevenly distributed throughout the planetesimal, leading to localized aqueoues alteration, rather than widespread. Consequently, it is important to have a better understanding of the geochemical and petrographic limits of the CO group, as meteorites such as MIL 07687 could have strong implications on our understanding of their parent body(ies).

**Acknowledgements**

We wish to thank Valérie Magnin and Alejandro Fernandez-Martinez (ISTerre) for having managed the TGA and IRS measurements. We also wish to thank the reviewers for their insightful and constructive comments. This work was funded by the Centre National d'Etudes Spatiales (CNES-France) and by the ERC grant SOLARYS ERC-CoG2017-771691.

**Table 1: Summary of the spectral data used for comparison with MIL 07687**

| | | transmission IR spectroscopy | | reflectance NIR spectroscopy |
|---|---|---|---|---|
| | | matrix fragments | bulk samples | bulk samples |
| | MIL 07687 | Prestgard and Bonal (2019) | this work | Eschrig et al. (2019a) |
| CO3 | DOM 08006 | Prestgard and Bonal (2019) | this work | Eschrig et al. (2019a) |
| | LAP 031117, DOM 03238 | Prestgard and Bonal (2019) | x | x |
| | ALHA 77003 | x | this work | Eschrig et al. (2019a) |
| | *NWA 8631, NWA 11889, El Médano 389* | x | this work | this work |
| | *MIL 090073* | x | x | this work |
| | MIL 03442, MIL 090785 | x | x | Beck et al., (2021) |
| CMs | *LEW 85311* | x | Beck and Garenne (2012) | This work |
| | QUE 97990 | x | Beck and Garenne (2012) | Potin et al. (2020) |
| | EET 87522 | x | Quirico et al., 2018 | This work |
| | NWA 11588 | Bonal, (unpublished) | x | x |
| CRs | MET 00426, Renazzo | Bonal et al. (2011a-e) | this work | x |
| | QUE 99177, | Bonal et al. (2011a-e) | x | This work |
| | Renazzo, QUE 99177 | Bonal et al. (2011a-e) | x | x |
| Ungrouped | EET 83355 | x | this work | this work |

DOM stands for Dominion Range, EET for Elephant Moraine, GRA for Graves Nunataks, LAP for LaPaz Icefield, MET for Meteorite Hills, NWA for North West Africa, and QUE for Queen Alexandra Range. The cross "x" means that no data were considered here for comparison with MIL 07687. Names in italic are peculiar CO-CM clan members, with properties perhaps intermediate between the CO and CM groups. Although currently listed as a CO3, MIL 090073 is likely a CM2 chondrite (Alexander et al., 2018).

Table 2: Summary of mass losses in the studied CO chondrites as a function of temperature by TGA.

| Sample | Class. | Petrographic grade | Weathering grade | TGA <200°C | TGA 200-350°C | TGA 350-700°C | TGA 200-700°C | TGA 77 |
|---|---|---|---|---|---|---|---|---|
| MIL 07687 | | 2 | Ce | 5.9 | 3.9 | 5.8 (8.9) | 9.7 (14.9) | 5.9 |
| DOM 08006-21 | CO | 3 | | 3.8 | 1.3 | 1.4 (6.6) | 2.7 (12.4) | 1.5 |
| DOM 08006-53 | CO | 3 | A/B | 3.4 | 1.4 | 1.5 (6.9) | 2.9 (13.2) | 1.7 |
| DOM 03238 | CO | 3.1 | B | 2.4 | 1.3 | 1.6 (5.7) | 2.9 (10.4) | 2.2 |
| MIL 05104 | CO | 3.1 | B | 2.2 | 1.4 | 1.4 (4.0) | 2.7 (8.0) | 1.6 |
| MIL 07193 | CO | 3.1 | A | 2.5 | 1.7 | 1.7 (5.0) | 3.4 (10.0) | 2.0 |
| MIL 05024 | CO | 3.1 | A/B | 2.2 | 1.6 | 1.4 (4.2) | 3.0 (8.8) | 1.8 |
| ALHA 77003 | CO | >3.6 | Ae | 1.9 | 1.2 | 1.3 (3.7) | 2.5 (7.3) | 1.4 |
| Moss | CO | >3.6 | Fall | 0.2 | 0.2 | n.d. | 0.2 (0.6) | n |

The petrographic grades of CO chondrites were extracted from Bonal et al. (2016) and Quirico et al. (2009). The values in parenthesis have been normalized according to the matrix abundance: 65% for MIL 07687, 22% for DOM 08006 (Davidson et al., 2019), 28% for DOM 03238 (Choe et al., 2010) and assumed 34% for all remaining CO3s (Weisberg et al., 2006).

Table 3: Petrographic properties of MIL 07687 compared to other CO-CM clan members

| | Matrix abundance (vol%) | Average chondrule size (μm) | Refractory inclusions | Bulk oxygen isotopic measurements | Classification (July, 2021) |
|---|---|---|---|---|---|
| MIL 07687 | 65% | 160±90 | Rare, especially AOAs | Yes | CO3/C2-ung |
| Mdaouer | 62% | 130±90 | n.a. | No | CO3.0 |
| Acfer 094** | 62.5% | 240 | CAIs and AOAs are present | Yes | C2-ung |
| EET 83226^ | n.a. | 120 | n.a. | Yes | C2-ung |
| El Médano 200 | 55% | 130±80 | Refractory | Yes (acid washed) | C3 |
| LEW 85311** | n.a. | 190 | Refractory inscusions present | Yes | CM2-an |
| NWA 11889 | 61% | 180±70 | n.a. | Yes (acid washed) | CO3.1-an |

Petrographic properties of MIL 07687 compared to those of average COs, CMs, the ungrouped meteorite Acfer 094, and a few unusual CO3s. Data associated with an asterisk (*) refer to values from Weisberg et al. (2006). Those with a double asterisk (**) were extracted from Choe et al. (2010). (^) refer to values extracted from Abreu et al. (2018). The MIL 07687 values refer to the information provided in this manuscript. All other information is from the Meteoritical society database. The average chondrule size of MIL 07687 (N= 49 measurements) was based on publicly available thin slice images at https://curator.jsc.nasa.gov/antmet/samples/petdes.cfm?sample=MIL07687

**Fig. 1:** Oxygen three-isotope graph showing bulk average O-isotopic measurements of MIL 07687 (black rhombus), relative to those of other CO-CM members. The "anomalous, unusual or atypical" chondrites correspond to meteorites with petrographic descriptions similar with MIL 07687, or with overlapping traits with the CO and CM groups, as well as CO-CM clan members currently listed as ungrouped in the Meteoritical society database (February, 2021). The "Primitive Asuka CMs" correspond to the CMs 2.8 – 3.0 studied by Kimura et al. (2020). The O-isotopic values are from Clayton and Mayeda (1999), Greenwood and Franchi (2004), Greenwood et al. (2018), Torrano et al. (2020) and the Meteoritical society database.

**Fig. 2:** Oxygen three-isotope graph showing bulk average O-isotopic measurements of MIL 07687 (black rhombus), relative to those of other CO-CM members. The "anomalous, unusual or atypical" members correspond to meteorites with petrographic descriptions similar with MIL 07687, or with overlapping traits with the CO and CM groups, as well as CO-CM clan members currently listed as ungrouped in the Meteoritical society database (February, 2021). The "Primitive Asuka CMs" correspond to the CMs 2.8 – 3.0 studied by Kimura et al. (2020). The O-isotopic values are from Clayton and Mayeda (1999), Greenwood and Franchi (2004), Alexander et al. (2018), Torrano et al. (2020), and the Meteoritical society database. The O-isotopic of Antarctic end-member minerals are also from Alexander et al. (2018).

**Fig. 3:** Transmission spectra of matrix fragments from MIL 07687 acquired under vacuum and at T = 250°C, but 300°C for fragments 3, 4 and 9. These have been baseline corrected and normalized according to the maximum intensity of their 10-μm (Si-O stretching) band. Notice the similarities between all twelve fragments.

**Fig. 4:** 3-μm IR transmission spectra of a representative MIL 07687 matrix fragments (#9) compared to those measured on matrix fragments of various CO3, CR2 and CM2 chondrites. All spectra were measured at under vacuum and at T = 300°C and have been normalized according the peak absorbance of their 10-μm bands. Vertical offsets have been added for the purpose of clarity. Petrographic grades are from Rubin et al. (2007), the meteoritical bulletin, Harju et al. (2014) and Bonal et al. (2016).

**Fig. 5:** Matrix IR transmission spectra of MIL 07687 showing evidence of carbonates (green lines). The brown line indicates sulphates and/or evaporites, consistent with the meteorite's extensively weathered nature.

**Fig. 6:** Normalized IR transmission spectra of bulk MIL 07687 compared to various type-2 CCs. Included are bulk spectra of enstatite, iron-poor olivine (Fo91) and iron-rich olivine (Fo66), for comparison. Vertical offsets have been added for the purpose of clarity. The dashed vertical lines correspond to signatures of enstatite (in brown), iron-poor (dark grey) and iron-rich (light grey) olivine. The petrographic grades for CMs are from Rubin et al. (2007), the meteoritical society database, Lee et al. (2019). Those for CR2s are from Harju et al. (2014).

**Fig. 7:** Normalized IR transmission spectra of bulk MIL 07687 compared to various CO3 chondrites. Included are bulk spectra of enstatite, iron-poor olivine (Fo91) and saponite, for comparison. Vertical offsets have been added for the purpose of clarity. The dashed vertical lines correspond to signatures of enstatite (in brown), iron-poor (dark grey), iron-rich olivine (light grey) and saponite (blue). The metamorphic grades were provided by the Meteoritical Society database and Bonal et al. (2016). The exception is El Médano 389, for which the metamorphic grade is not yet published.

**Fig. 8:** Normalized Vis-NIR reflectance spectrum of MIL 07687 compared to those measured on various type 2 CCs (left) and CO3s (right). NWA 8631 and NWA 11889 are matrix-enriched CO3s. Most CO3s chosen for this plot are highly weathered (the exception is DOM 08006), to better compare with MIL 07073. MIL 090073 is still recognized as a CO3 in the Meteoritical society database, but is a likely heated CM2 (Alexander et al. 2018). Vertical offsets have been added for the purpose of clarity.The asterisks (*) indicate samples measured under ambient temperature. MIL 03442 and 090785 were measured under ambient conditions. "^" indicate metamorphic grades extracted from the Meteoritical bulletin. The metamorphic grades of MIL 03442 and 090785 are assumed based on their parity with MIL 05042 (CO3.1: Bonal et al., 2016). The spectra were normalized according to their reflectance at 0.56 μm, with the exception of MIL 03442 and 09785 (0.6 μm).

**Fig. 9:** Normalized Vis-NIR reflectance spectrum of MIL 07687 compared to those measured on various type 2 CCs and CO3s. Vertical offsets have been added for the purpose of clarity. Both MIL 07687 was measured under vacuum at ambient temperature. MIL 03442 and 090785 were measured under ambient conditions. The spectra were normalized according to their reflectance at 0.56 μm.

**Fig. 10**: TGA and DTG curves of MIL 07687 and all CO3 samples measured in this work. The spectra are color-coded according to metamorphic grade, although note that MIL 05104, MIL 05024 and MIL 07193 are most likely

paired. The profile of the mass loss curve of MIL 07687 is close to those observed in CO3s. The DOM 08008-21 and 53 samples correspond to the 30 mg and 50 mg DOM 08006 samples respectively.

**Fig. 11**: DTG curves of MIL 07687 and all CO3 samples (except DOM 08006-53) measured in this work, with the annotated minerals contributing to the data, based on the interpretations of this work. The spectra are color-coded according to metamorphic grade. Note that the $650 - 900°C$ region may also be contributed by the reduction of magnetite by organic matter, especially in CO3s.

**Fig. 12:** Summed intensity (integrated absorbance) of the 4.6 and 4.75-µm bands combined, as a function of temperature (represented as time) in the IRS spectra. The spectra are color-coded according to metamorphic grade. The sample mass is the same for each chondrite allowing a direct comparison of the band intensity. Variability prior to 3000s is most likely related to contamination from the nearby carbon dioxide bands. The IRS spectra are plotted according to the time, rather than temperature. This is because the spectra were 180 seconds prior to the increments in temperature.

**Fig 13:** Normalized (baseline corrected) IRS 4.6-µm and 4.75-µm bands of MIL 07687 in comparison to CO3 chondrites. The colors are representative of the metamorphic grade attributed by Bonal et al. (2007, 2016) and Quirico (2009): Purple (3.00) – Red (>3.6). Notice the vibrational complexity of the 4.75-µm bands. Only the spectra obtained from 30 mg samples are compared here (thus not including our measurements of DOM 08006-53).

**Fig. 14:** Comparison of Raman spectra acquired on the matrices of MIL 07687 (from Bonal et al. 2016) and type 2 and type 3 chondrites. (a) Average and G-band normalized Raman spectra of MIL 07687 (n = 20) in comparison to Mukundpura (CM2; n = 35) and DOM 08006 (CO 3.0; n = 34). (b) G-band parameters (Full Width at Half Maximum of the G-band as a function of its position). (c) D-band parameters (Full Width at Half Maximum of the D-band as a function of its position). Data for type 2 chondrites are from Quirico et al. (2018) and type 3 chondrites from Bonal et al. (2016). "R1, "R2" and "R3" type 2 chondrites are more or less heated (see Quirico et al., 2018 for details).

**Fig. 1:**

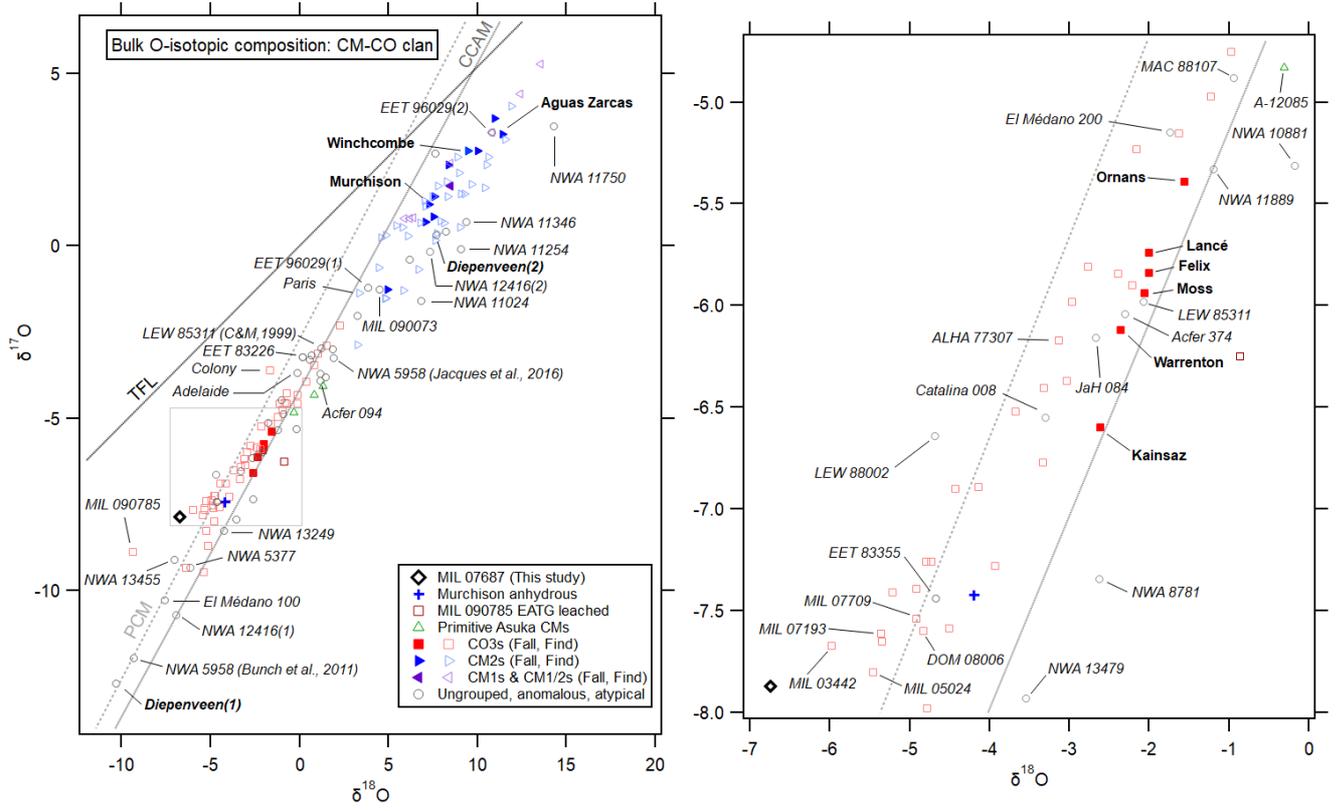

5 **Fig. 2:**

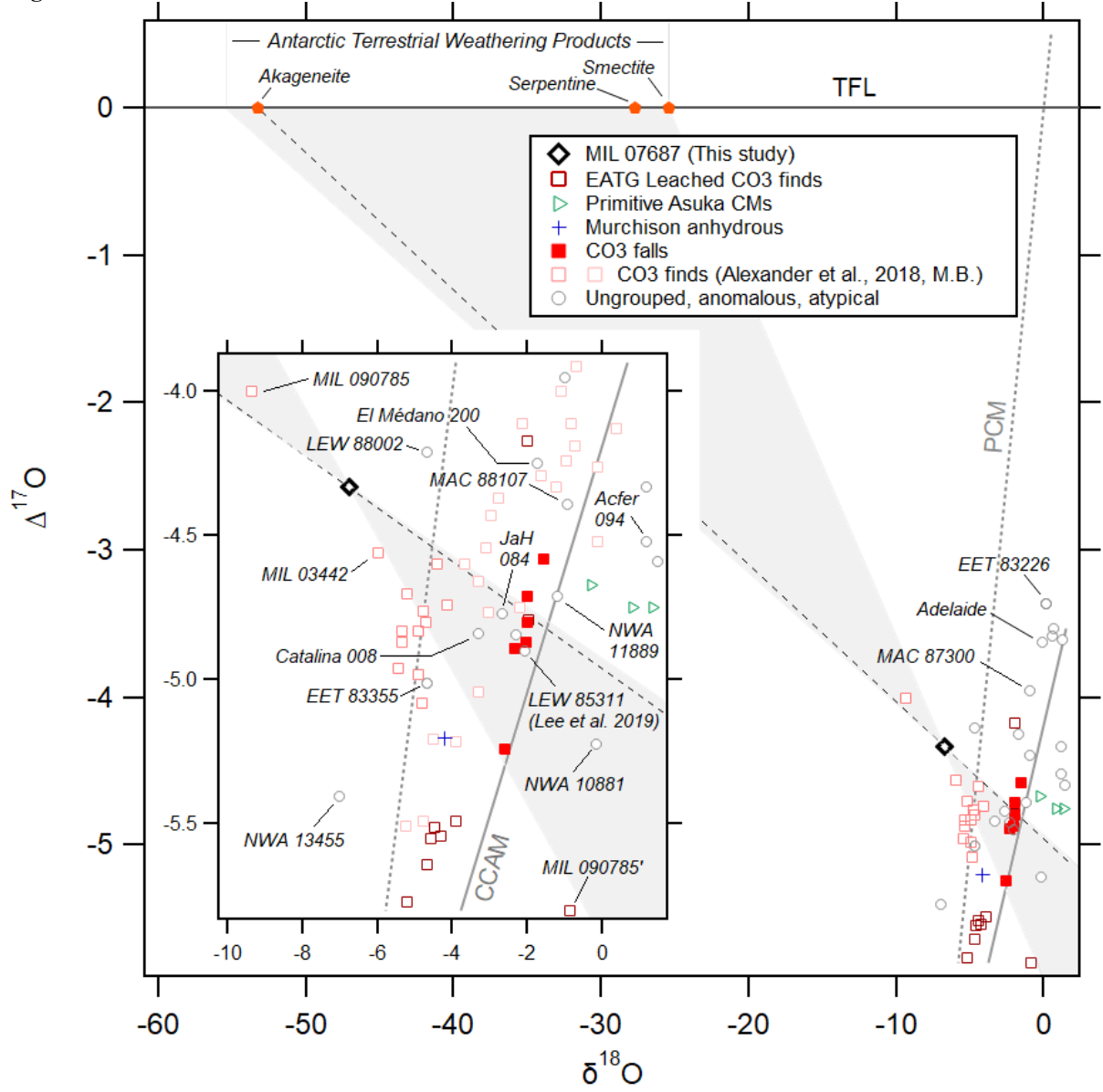

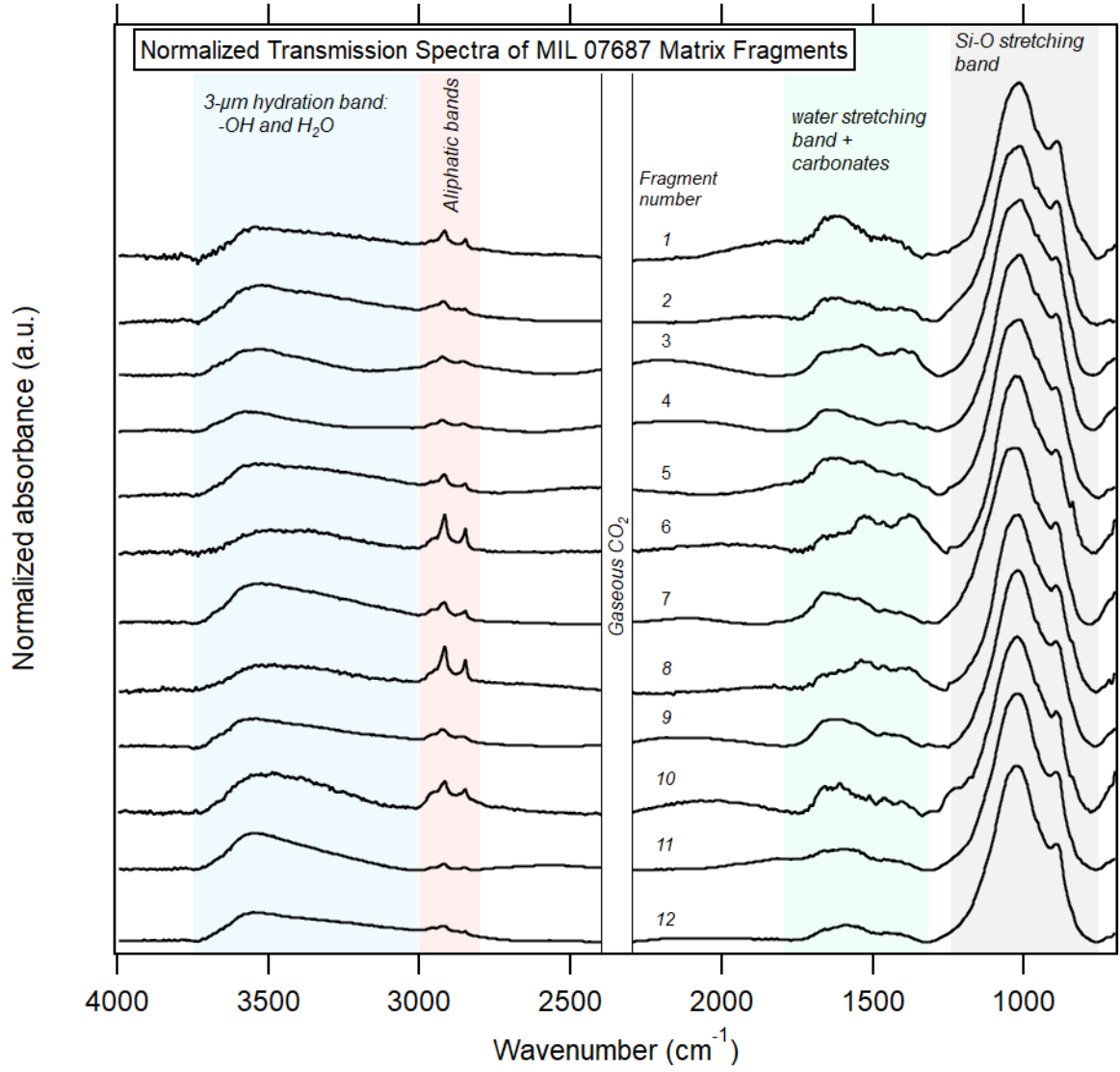

10    **Fig. 4:**

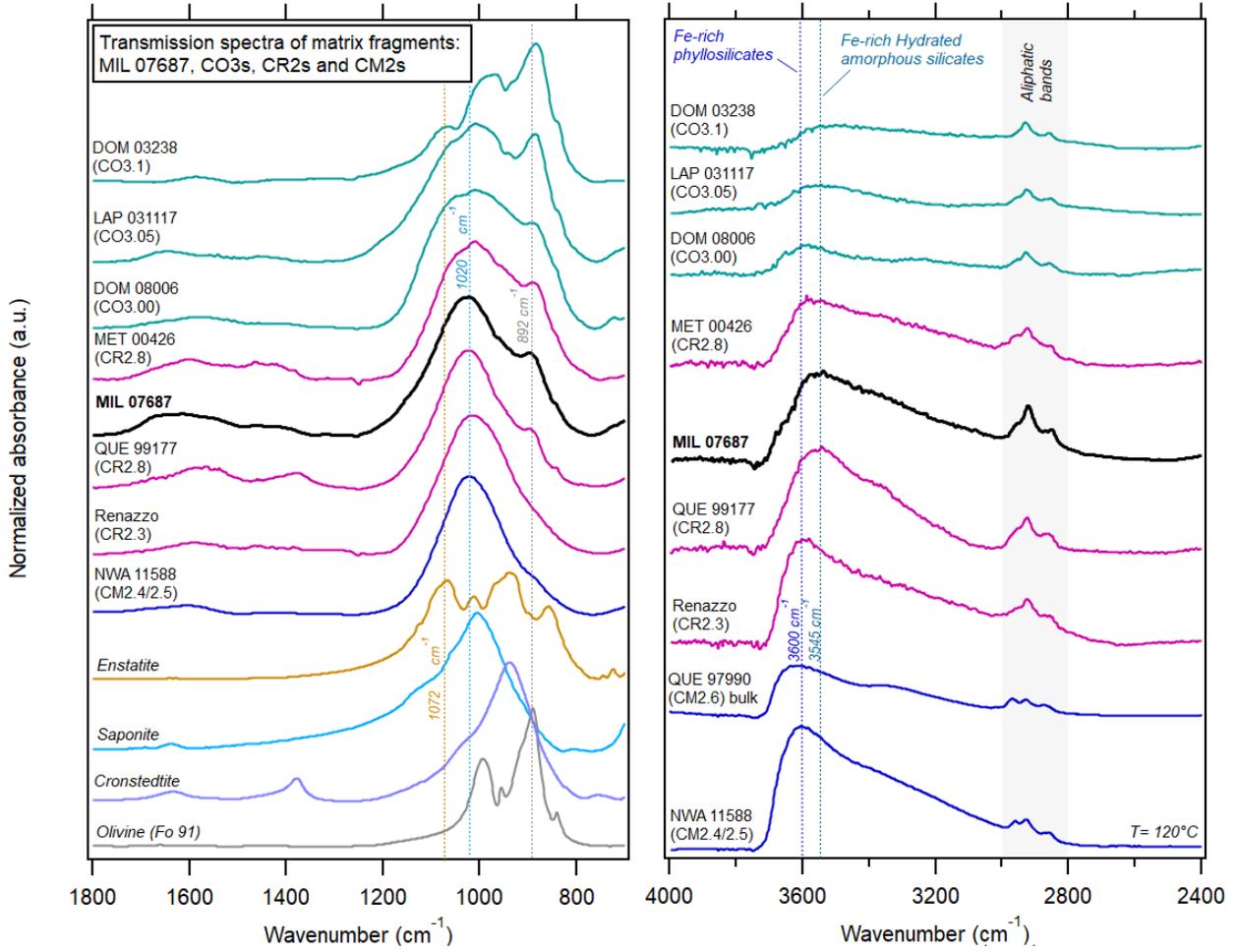

11
12
13    **Fig. 5:**

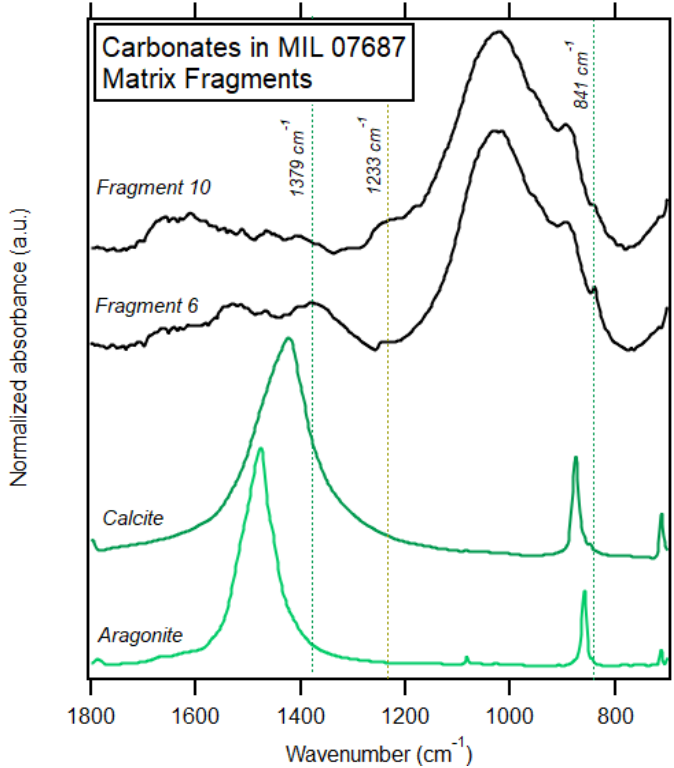

14
15

**Fig. 6:**

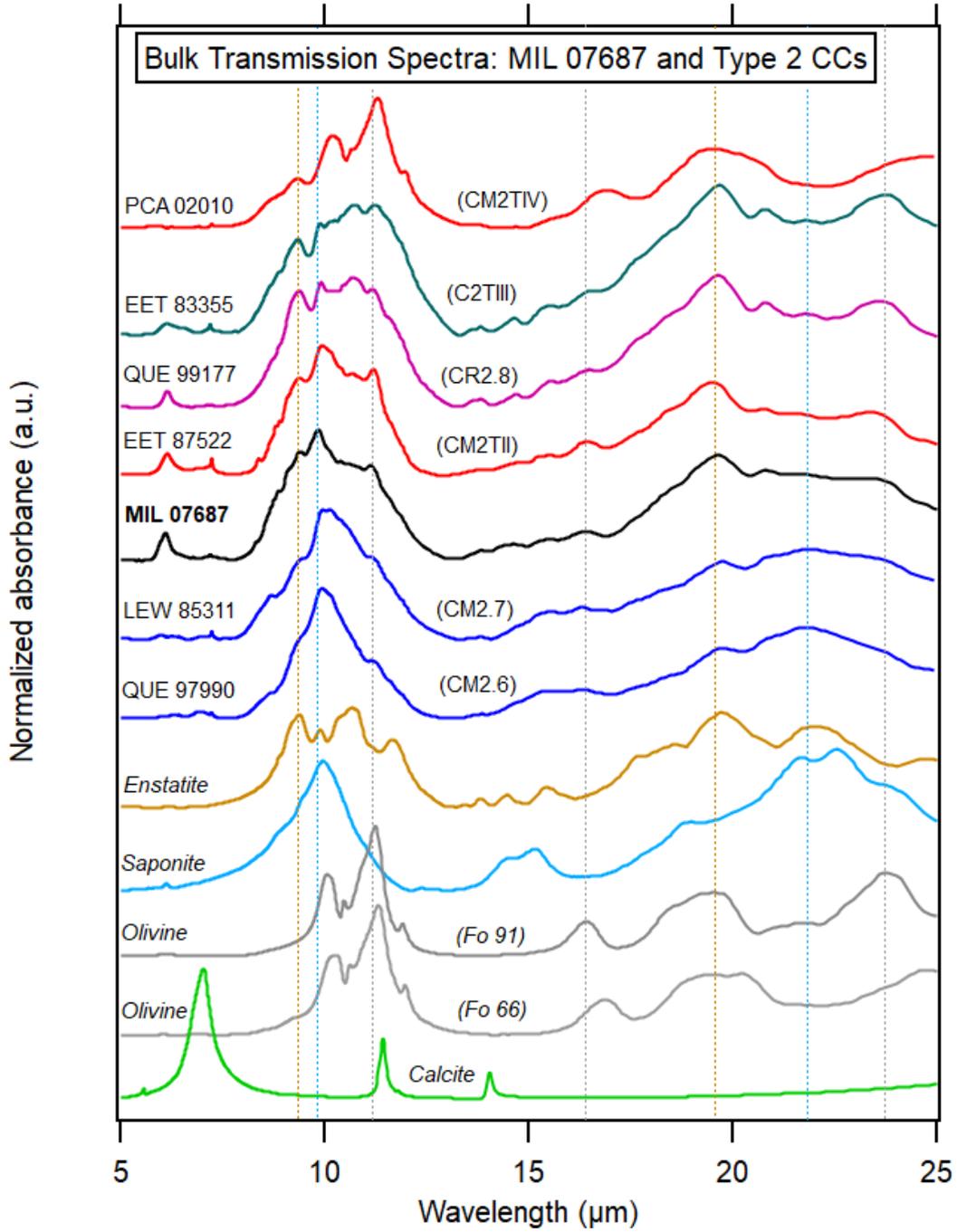

Bulk Transmission Spectra: MIL 07687 and Type 2 CCs

PCA 02010 (CM2TIV)
EET 83355 (C2TIII)
QUE 99177 (CR2.8)
EET 87522 (CM2TII)
MIL 07687
LEW 85311 (CM2.7)
QUE 97990 (CM2.6)
Enstatite
Saponite
Olivine (Fo 91)
Olivine (Fo 66)
Calcite

Normalized absorbance (a.u.)

Wavelength (μm)

**Fig. 7:**

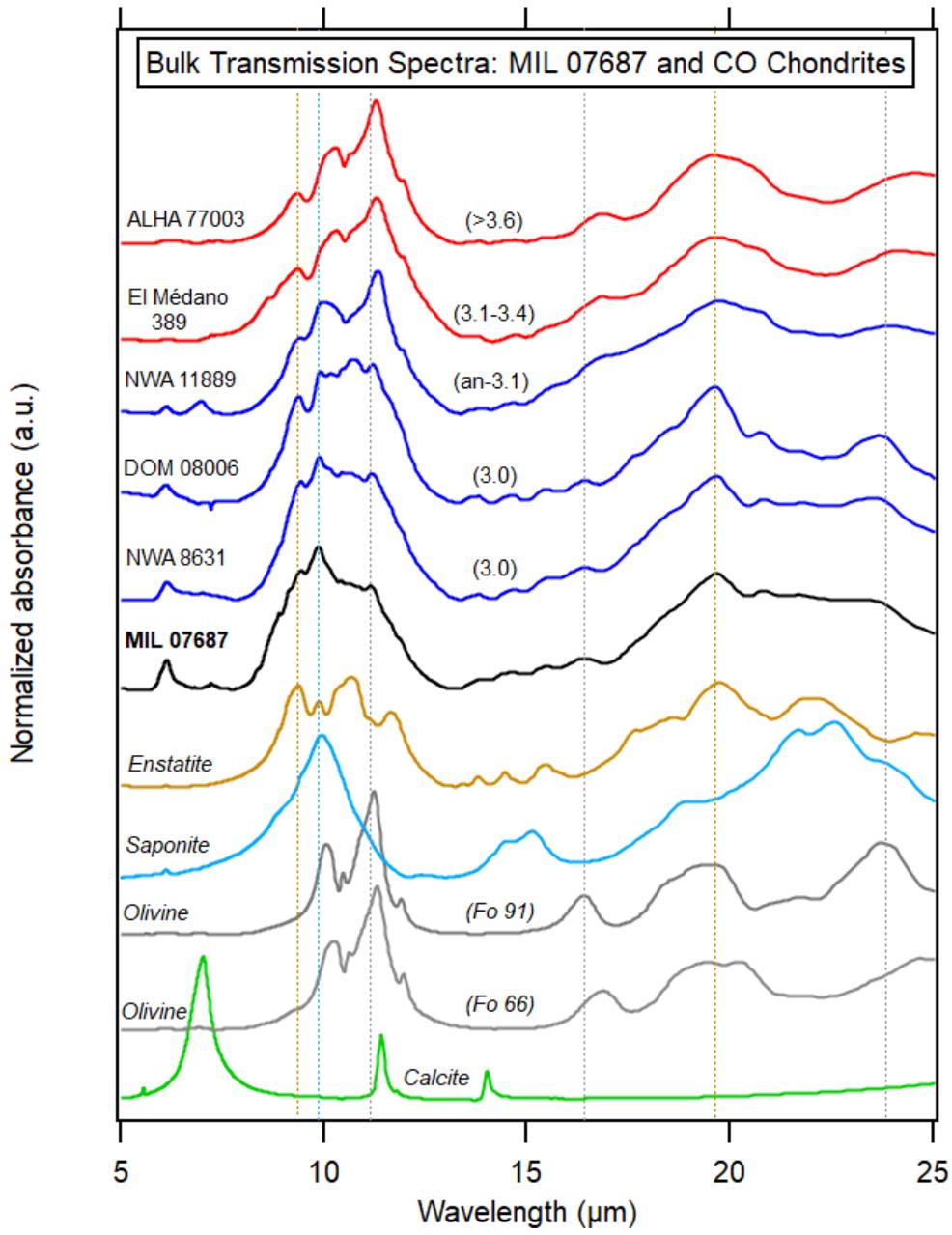

**Fig 8:**

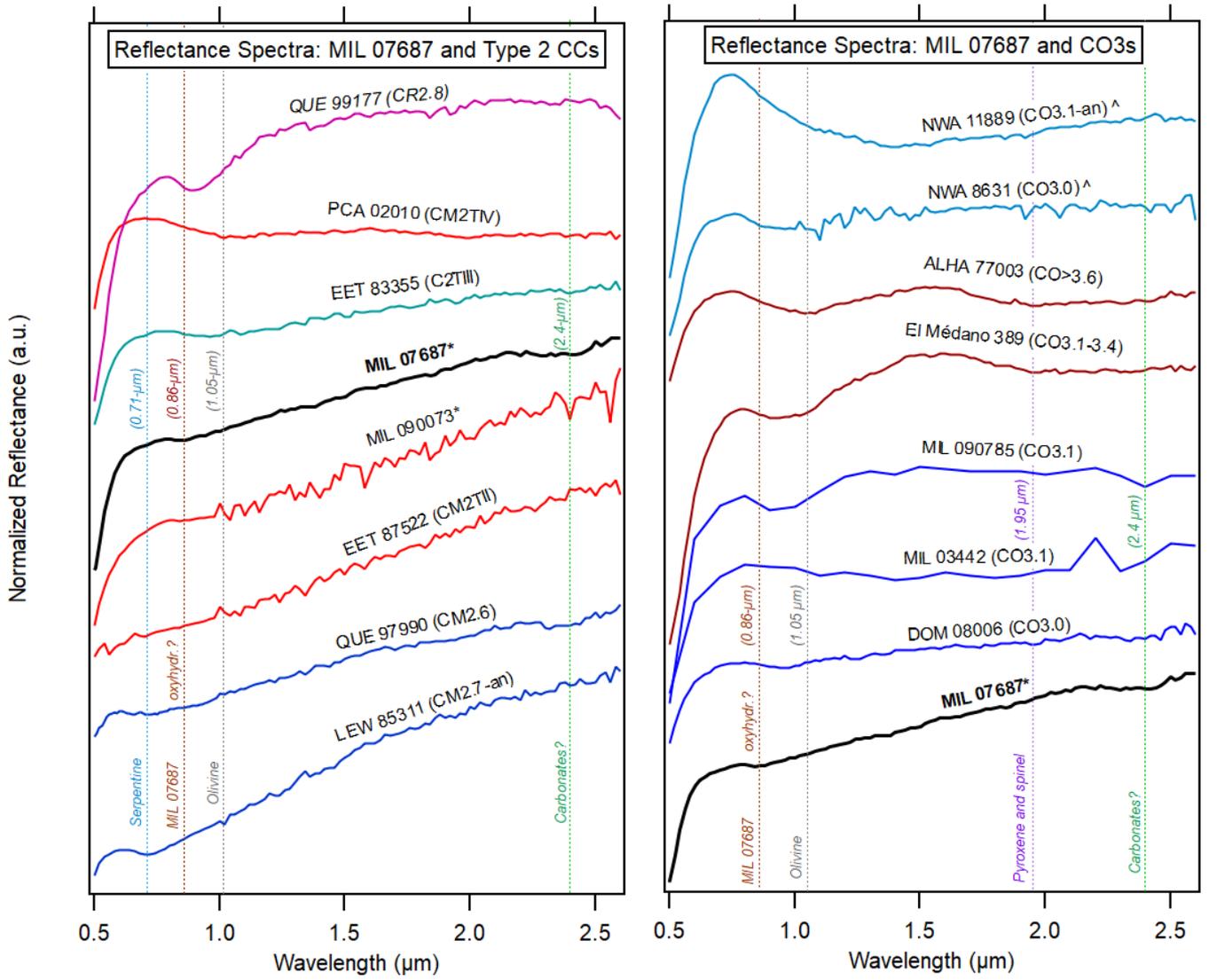

23    **Fig 9:**
24

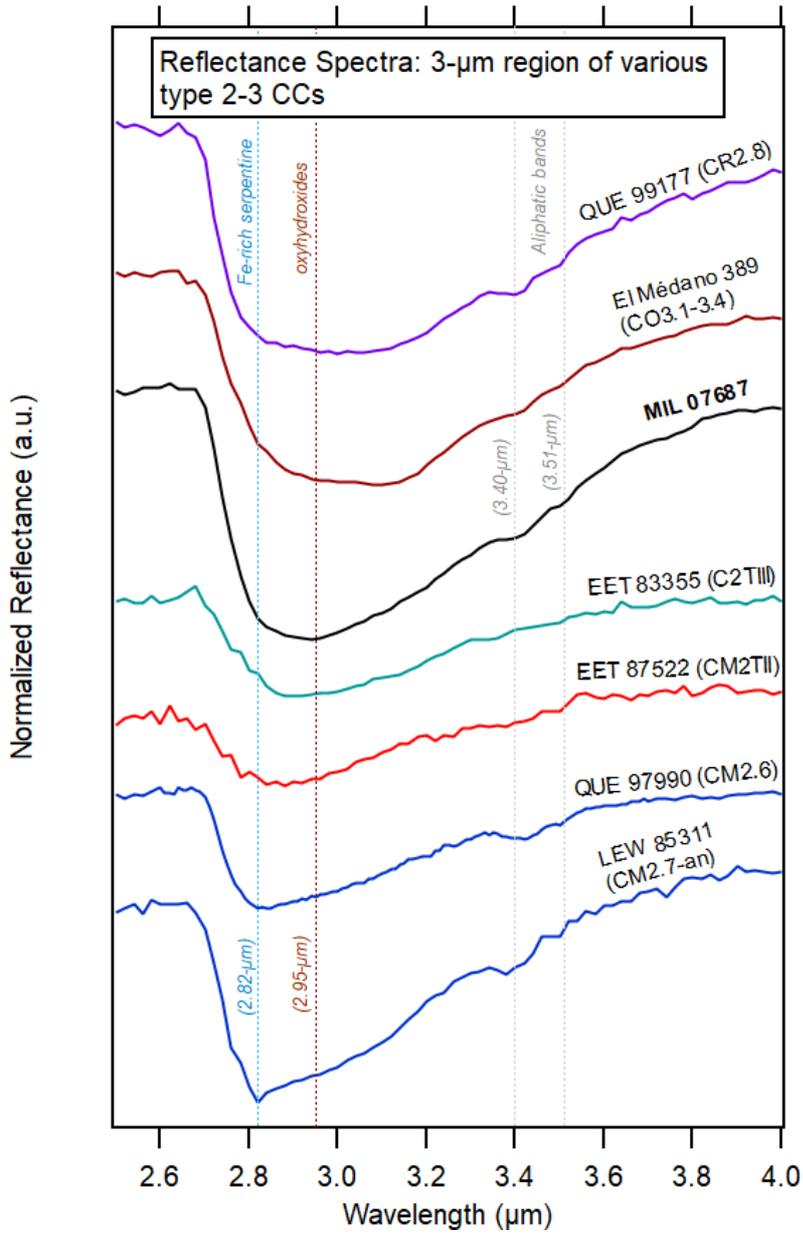

**Fig. 10:**

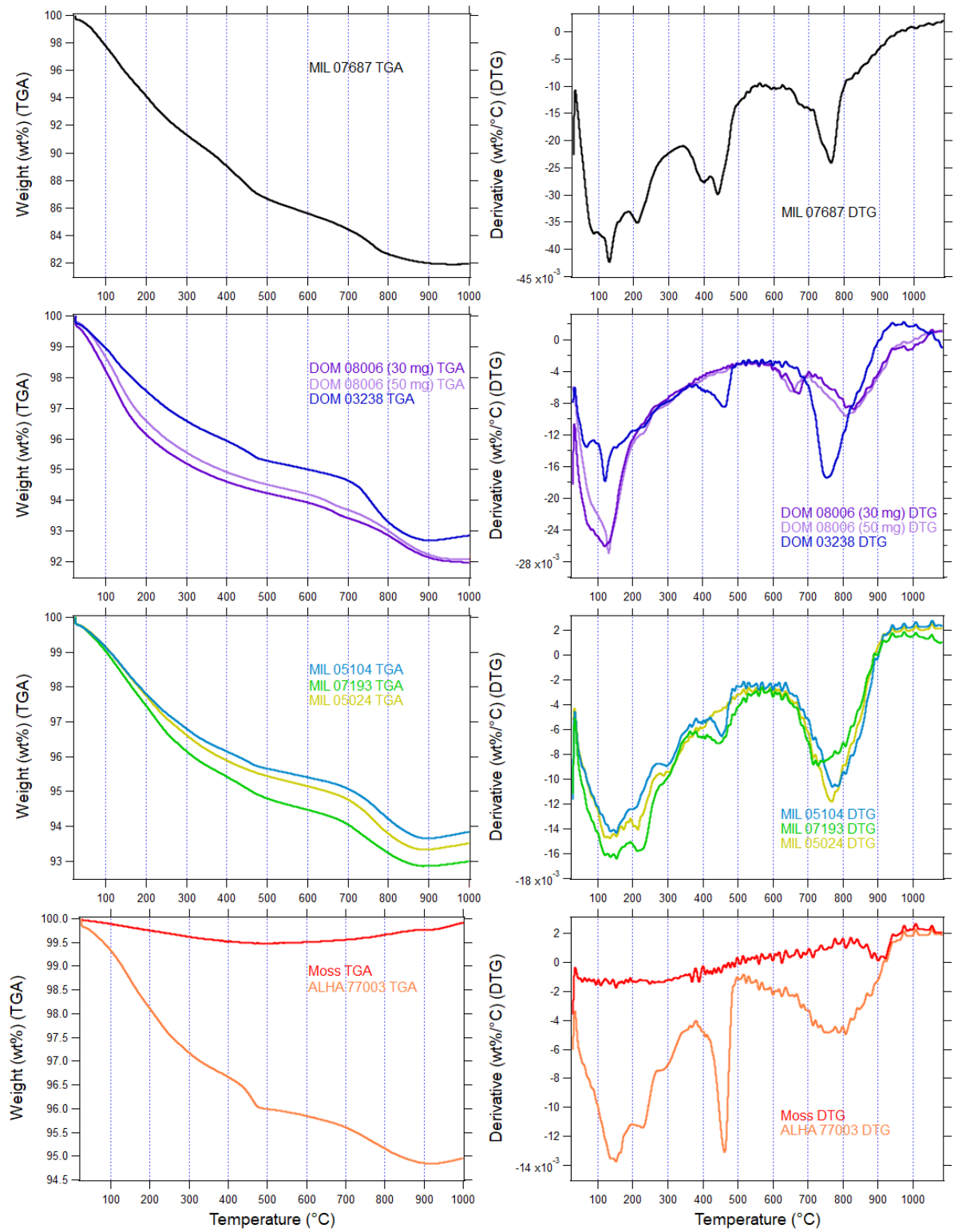

**Fig. 11:**

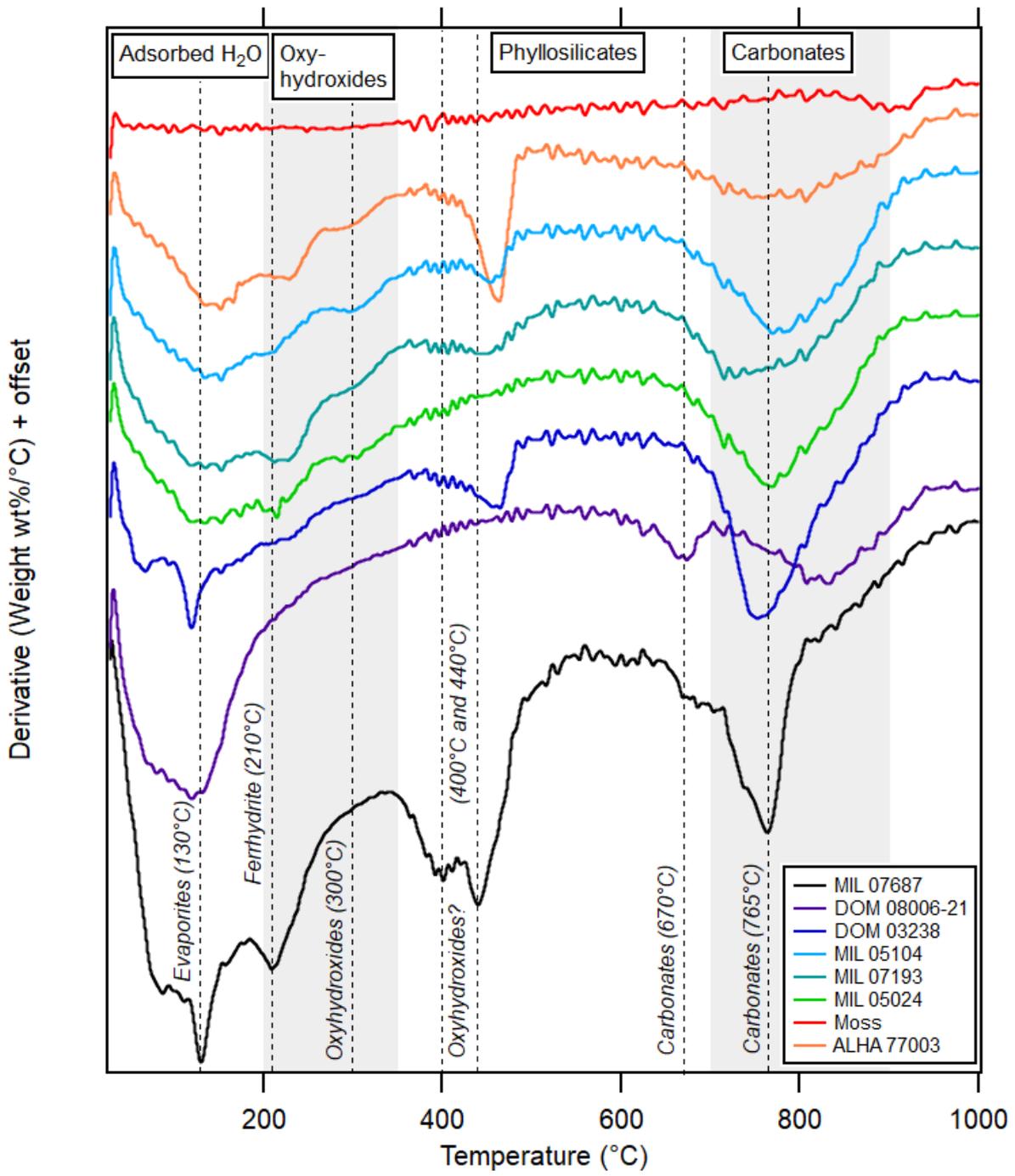

39    **Fig. 12:**

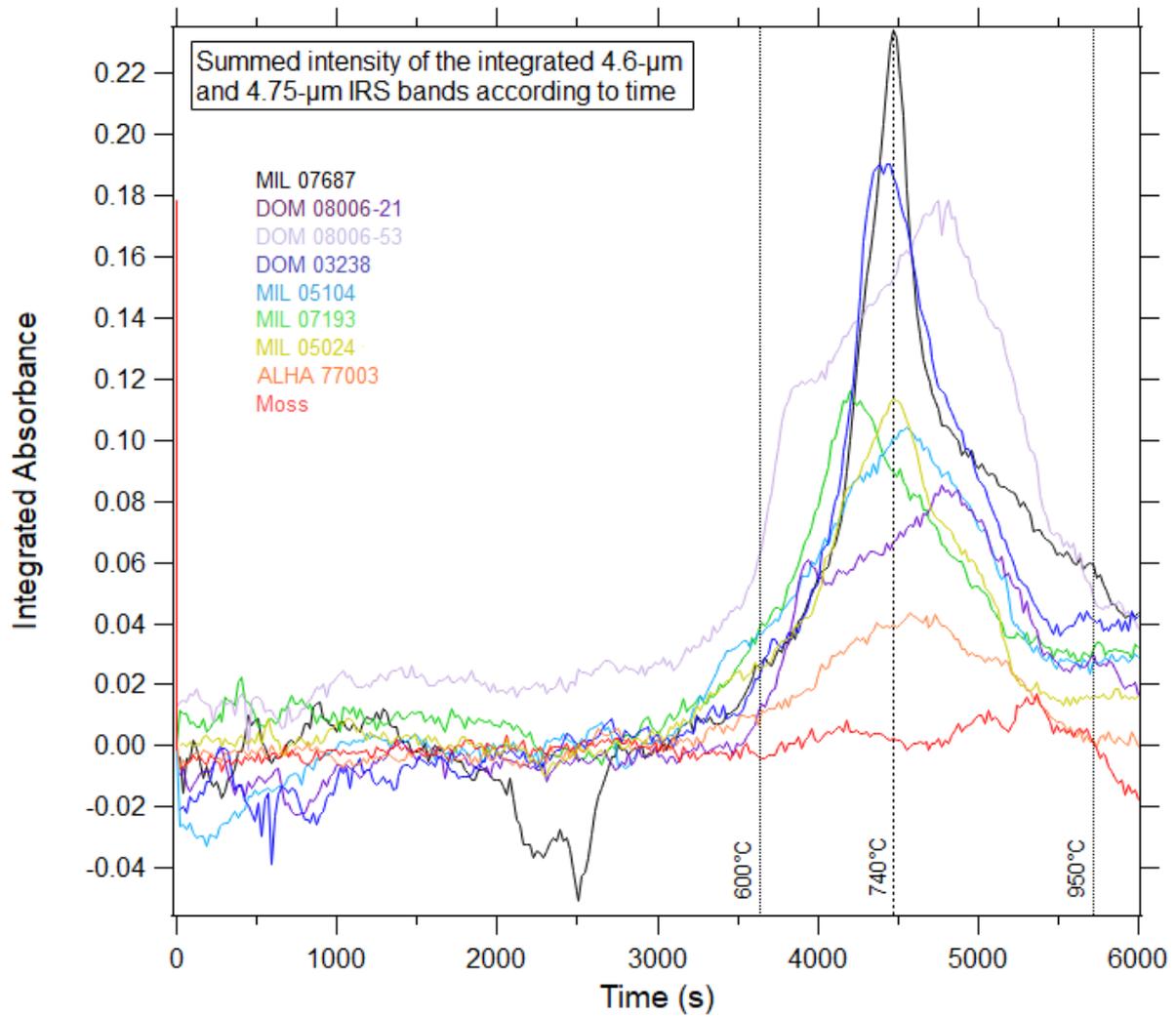

Summed intensity of the integrated 4.6-μm and 4.75-μm IRS bands according to time

MIL 07687
DOM 08006-21
DOM 08006-53
DOM 03238
MIL 05104
MIL 07193
MIL 05024
ALHA 77003
Moss

40
41
42
43
44
45
46
47
48

**Fig. 13**:

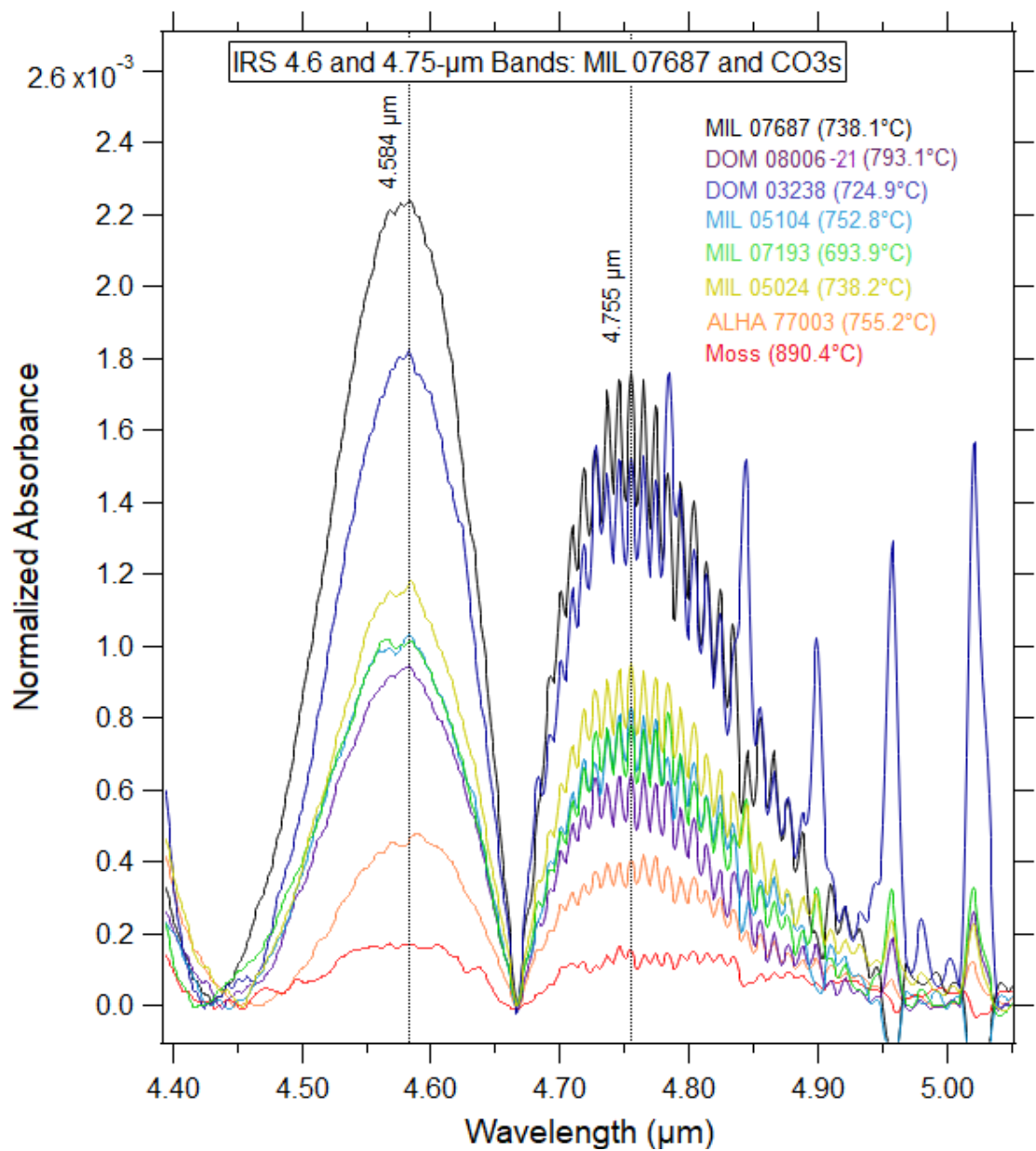

IRS 4.6 and 4.75-µm Bands: MIL 07687 and CO3s

MIL 07687 (738.1°C)
DOM 08006-21 (793.1°C)
DOM 03238 (724.9°C)
MIL 05104 (752.8°C)
MIL 07193 (693.9°C)
MIL 05024 (738.2°C)
ALHA 77003 (755.2°C)
Moss (890.4°C)

4.584 µm

4.755 µm

Normalized Absorbance

Wavelength (µm)

54    **Fig. 14:**

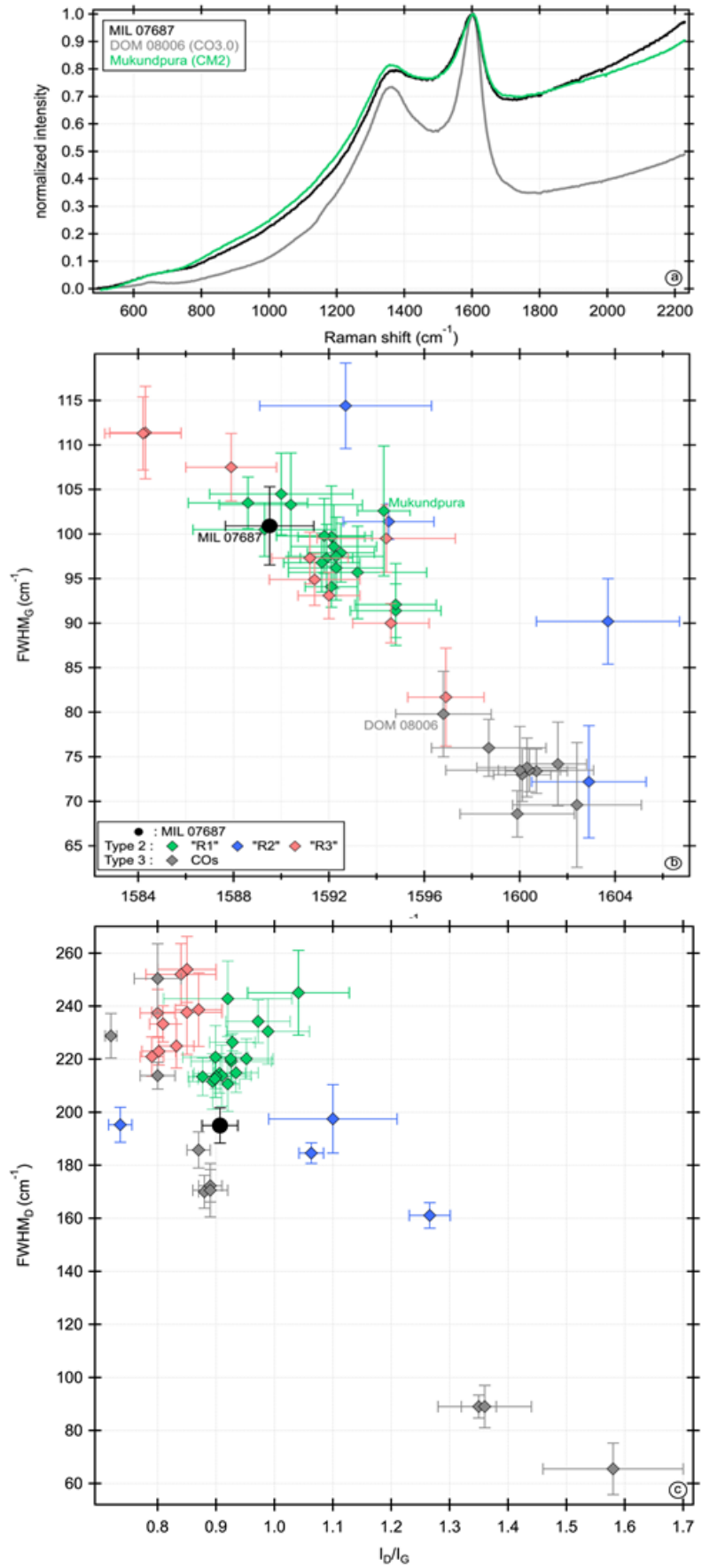

55